\documentclass[twocolumn,aps]{revtex4-1}

\usepackage{graphicx}
\usepackage{amssymb}
\usepackage{amsfonts}
\usepackage{amsmath}
\usepackage{amsbsy}
\usepackage{natbib}
\usepackage{comment}

\usepackage{color}
\usepackage{float}

\renewcommand{\vec}[1]{\mbox{\boldmath$\mathrm{#1}$}}
\newcommand{\be}{\begin{equation}}
\newcommand{\ee}{\end{equation}}
\newcommand{\ben}{\begin{eqnarray}}
\newcommand{\een}{\end{eqnarray}}

\begin{document}

\title{Thermoelastic enhancement of the magnonic  spin Seebeck effect in thin films and bulk samples}

\author{L. Chotorlishvili$^1$, X.-G. Wang$^{2}$,  Z. Toklikishvili$^3$,  J. Berakdar$^1$}

\address{$^1$ Institut f\"ur Physik, Martin-Luther Universit\"at Halle-Wittenberg, D-06120 Halle/Saale, Germany \\
$^2$ School of Physics and Electronics, Central South University, Changsha 410083, China \\	
$^3$  Faculty of Mathematics and Natural Sciences, Tbilisi State University, Chavchavadze av.3, 0128 Tbilisi, Georgia}

\begin{abstract}

A non-uniform temperature profile may generate  a pure spin current in magnetic films, as observed for instance in the spin Seebeck effect.
In addition,  thermally induced elastic  deformations may set in that could affect the spin current. A
self-consistent theory of the magnonic spin Seebeck effect  including  thermally activated magneto-elastic effects is presented and  analytical expressions for the thermally activated deformation tensor and dispersion
relations for coupled magneto-elastic modes are obtained. We derived analytical results for   bulk (3D) systems and  thin magnetic (2D) films.
 We observed that the displacement vector and the deformation tensor in  bulk systems decay asymptotically
as $u\sim1/R^{2}$ and $\varepsilon\sim1/R^{3}$, respectively, while the decays in  thin magnetic films proceed slower following $u\sim1/R$ and $\varepsilon\sim1/R^{2}$.
The dispersion relations evidence a strong anisotropy in the magnetic excitations.
We observed that a  thermoelastic steady state deformation may lead to
 both an enchantment or a reduction of the gap in the magnonic spectrum.
The reduction of the gap increases the number of magnons contributing to the spin Seebeck effect
and offers new possibilities for the thermoelastic control of the Spin Seebeck effect.

\end{abstract}

\date{\today}

\maketitle

\section{Introduction}
By virtue of magnetoelastic coupling elastic deformations may trigger a magnetization dynamics and (magneto)elastic waves maybe launched due to spin motion. The study of
 elastically activated magnetic dynamics in ferro- and antiferromagnetic materials dates back to the late 1950s  starting  with seminal independent works  by A. I. Akhlezer, V. G. Ber\'yakhtar, S.V. Peletminsky \cite{Akhlezer} and C. Kittel \cite{Kittel}. Further imputes came from the discovery  of the magnetoelastic-gap \cite{Borovik-Romanov,Tasaki,Turov}  that  bears some resemblance to  spontaneous symmetry breaking  \cite{Turov2}.
Since the magnetically excited elastic waves affect in turn the magnetization dynamics the established magnetoelastic gap,  being    a second order effect, is proportional to the square of the magnetoelastic coupling constant.
Thus, the magnetoelastic gap is usually  quite small compared to the gap in the magnonic spectrum which is induced for instance   by a magnetocrystalline anisotropy or by external field terms. \\
A thermal heating leading to a steady state elastic  deformation may serve as an alternative  for activating (magneto) elastic modes that occur in ferromagnetic  films and heterstructures  \cite{exp1,exp2,exp3,exp4,exp5,exp6,exp7,theo1,theo2,theo3,theo4}.
 Elasticity involving  non-isothermal deformations is part of  the well-established field of
  thermoelasticity \cite{Kupradze,Landau}.  An important question in the context of the present paper is to which
  extent a steady state thermoelastic deformation  influences the magnetoacoustic effect. Due to the grossly  different  time scales of the dynamics, a steady state thermoelastic deformation is swiftly established (meaning equilibrated with the external thermal bath), and is basically unaffected   by  the much slower magnetization dynamics. The
magnetization dynamics  may well be sensitive to  thermoelastic deformation, however. We will investigate here the theoretical aspects of  thermal magnetoacoustics, i.e. thermally activated magnetoelastic effects with a special focus on phenomena of
interest to the active field of  spin caloritronics
\cite{Bauer,Barker,Lefkidis,Basso,Uchida,Kehlberger,Schreier,Basso2,Ritzmann,Kikkawa,JiUp13,EtCh14,khomeriki,Sukhov,Barnas,Saitoh,Weiler}.
Experimentally, the utilization of elasticity to steer  the magnetic dynamics is meanwhile accessible in a variety of setting.  For instance,
Rayleigh surface acoustic waves that may couple to spin ordering can be generated by irradiation with  laser pulses \cite{Crimmins}. This process may well be accompanied by local heating spreading away from the laser spot which in turn may launch temporally a  spin Seebeck  current.
Heating by laser pulses was employed for  experiments concerning the  time resolved spin Seebeck effect  \cite{Boona}.
Simulations for  time resolved spin Seebeck effect were presented in Ref. \cite{Etesami}. \\
A comprehensible study of the thermal magnetoacoustic effect should encompass  both, heating and elasticity aspects. Heating, for instance by  laser pulses leads to  a buildup of a nonuniform temperature distribution and possibly a temporal magnonic spin Seebeck effect.  Non-isothermal deformations may also contribute to magnetoelastic  activation of magnonic spin current. For example,
considering that non-isothermal deformation of the thin film may reduce gap in the magnonic spectrum,  the spin Seebeck effect may well be modified, for a reduction in the magnonic gap increases the number of magnons contributing to the spin Seebeck effect. In what follows we explore the link between the nonisothermal deformation ($\vec{R}$) tensor $\vec{\varepsilon}_{\xi \zeta}\big(\vec{R}\big)$ and the magnonic energy spectrum $\omega^{2}\big(q,\vec{R}, \vec{\varepsilon}_{\xi \zeta}\big(\vec{R}\big)\big)$ at the wave vector $q$.
We derive analytical solutions for the deformation tensor and implement it for the thermally activated magnetoelastic dynamics.
We analyze in details the 3D case of a Bulk sample and compare with a 2D case of a thin film.
Analytical results are complemented by full numerical micro-magnetic simulations. \\
The paper is organaized as follows: in  section II we introduce the model, in the section III we discuss the generalities of the magneto-thermal effect and derive explicit analytical expressions
for the displacement vector and for the  deformation tensor for local and non-local heat sources, section IV is dedicated to the dispersion relations for thermally excited magneto-elastic magnonic modes.
In  section V we present analytical results for the spin wave dispersion in thin films,  and in  section VI we analyze the results of the micromagnetic numerical calculations followed by
a summary and conclusions.

\section{General formulation}
We study the transversal magnetic  dynamics of a magnetoelastically coupled system as it described by the deformation-dependent time evolution of the unit vector field $\vec{M}$.
 We will work along a Landau-Ginzburg approach starting from the energy functional
\begin{eqnarray}
\label{Hamiltonian}
H=H_{m}+U_{mel}(\vec{R}).
\end{eqnarray}
The magnetic part $H_{m}$  can be broken down essentially into the exchange, magnetocrystalline anisotropy, and Zeeman terms, respectively (summation over repeated indexes is assumed)
\begin{eqnarray}
\label{magneticHamiltonian}
H_{m}=A_{\xi\zeta}\frac{\partial \vec{M}}{\partial x_{\xi}}\frac{\partial \vec{M}}{\partial x_{\zeta}}+K_{\xi\zeta}\vec{M}_{\xi}\vec{M}_{\zeta}-\vec{B} \cdot \vec{M},
\end{eqnarray}
where $A_{\xi\zeta}$ is the exchange stiffness, $K_{\xi\zeta}$ quantifies the magnetocrystalline anisotropy energy contribution, and $\vec{B}$ is an external magnetic field.
The magnetoacoustic energy density $U_{mel}(\vec{\vec{R}})$ reads
\begin{eqnarray}
\label{magnetoacoustic energ}
U_{mel}(\vec{\vec{R}})=\frac{B_{1}}{M_{s}^{2}}
M^{2}_{\xi}\varepsilon_{\xi \xi}+\frac{B_{2}}{M_{s}^{2}}
M_{\xi}M_{\zeta}\varepsilon_{\xi\zeta} .
\end{eqnarray}
Here $M_{S}$ is the saturation magnetization, $M_{\xi}(\vec{R}),~M_{\zeta}(\vec{R})$ are the
magnetization components along  the $\xi,\zeta=x,y,z$ axes, and $B_{1},~B_{2}$ are the magnetoelastic constants. The deformation tensor  has the explicit form
\begin{eqnarray}
\label{deformation tensor}
\varepsilon_{\xi\zeta}(\vec{R})=\frac{1}{2}\bigg(\frac{\partial u_{\xi}(\vec{R})}{\partial x_{\zeta}}+\frac{\partial u_{\zeta}(\vec{R})}{\partial x_{\xi}}\bigg),
\end{eqnarray}
where $u_{\xi}(\vec{R})$ is the component of the displacement vector.  The stress tensor of the system $\sigma_{\xi\zeta}$  satisfies the relation  $F_{\xi}=\frac{\partial \sigma_{\xi\zeta}}{\partial x_{\zeta}}$, where  $F_{\xi}$ is the component of the external force which is applied on the system.
In the absence of an external forces, equilibrium requires that $\frac{\partial \sigma_{\xi\zeta}}{\partial x_{\zeta}}=0$.  The stress and the deformation tensors are interrelated  via the algebraic relation
\begin{eqnarray}
\label{Stress and deformation}
\sigma_{\xi\zeta}=\frac{E}{1+\sigma}\bigg(\varepsilon_{\xi\zeta}+\frac{\sigma}{1-2\sigma}\varepsilon_{\xi\xi}\delta_{_{\xi\zeta}}\bigg).
\end{eqnarray}
Here $E$ is the elasticity modulus  and $\sigma$ is  Poisson's constant.
\section{Magneto-thermal effects in the 3D bulk system}
We will be dealing with small amplitude displacements in the 3D bulk system. Proceeding in a standard way, the
equation of motion for  elastic waves without an applied thermal bias follows as \cite{Landau}
\begin{eqnarray}
\label{elastic wave}
&&\rho \frac{d^{2}\vec{u}}{dt^{2}}=\frac{E}{2(1+\sigma)}\triangle \vec{u}+\nonumber\\
&&\frac{E}{2(1+\sigma)(1-2\sigma)} \nabla (\nabla \cdot \vec{u}).
\end{eqnarray}
In the presence of an applied thermal bias $\nabla T$, one derives the equation of motion for the thermo-elastic waves by adding the temperature term,
\begin{eqnarray}
\label{thermal elastic wave}
&&\rho \frac{d^{2}\vec{u}}{dt^{2}}=\frac{E}{2(1+\sigma)}\triangle \vec{u}+\nonumber\\
&&\frac{E}{2(1+\sigma)(1-2\sigma)}\nabla\big(\nabla \cdot  \vec{u}\big)-\frac{E\kappa \nabla T}{3(1-2\sigma)}.
\end{eqnarray}
 $\kappa$ is the thermal expansion coefficient, and $\nabla T$ is a temperature gradient  which is due to a laser heating, for instance.
Eq. (\ref{thermal elastic wave}) describes the dynamics of the elastic modes coupled to the magnetization dynamics via magnetoelastic coupling
 $U_{mel}(\vec{R})$ (see Eq. (\ref{magnetoacoustic energ})).  Classically, the magnetization dynamics follows the stochastic Landau-Lifshitz-Gilbert equation
\begin{eqnarray}
\label{LLG}
&&\frac{\partial\vec{M}(\vec{R},t)}{\partial t}=-\gamma \vec{M}(\vec{R},t)\times\bigg(\vec{H}_{eff}+\vec{h}\bigg)-\\
&&-\frac{A}{\hbar M_{S}}\vec{M}(\vec{R},t)\times\nabla^2\vec{M}(\vec{R},t)+\frac{\alpha}{M_{s}}\vec{M}(\vec{R},t)\times\frac{\partial\vec{M}(\vec{R},t)}{\partial t}\nonumber
\end{eqnarray}
with  the deterministic effective field  $\vec{H}^{\mathrm{eff}}=-\frac{\delta \vec{H}}{\delta \vec{M}}$  and
 $H=H_{m}+U_{mel}(\vec{R})$ complemented by  a random field $h(\vec{R},t)$ due to a  Gaussian white noise with the autocorrelation function
\begin{equation}
\label{white-noise}
     \begin{array}{l}
       \displaystyle{\left<h_{i}(t,\vec{R})h_{j}(t' ,\vec{R'})\right> =}\\
       \displaystyle{\frac{2k_{B}T\big(\vec{R}\big)\alpha}{\gamma M_{\mathrm{S}}a^3}
       \delta_{ij} \delta\big(\vec{R}-\vec{R'}\big) \delta\big(t-t'\big).}
     \end{array}
\end{equation}
Here $\alpha$ is the Gilbert damping, $\gamma=1.76 \cdot 10^{11}$~[1/(Ts)] is the gyromagnetic ratio,  $T\big(\vec{R}\big)$ is the local temperature formed in the system
and $M_{\mathrm{S}}$ is the saturation magnetization. The
magnonic spin current tensor is evaluated as
\begin{eqnarray}
\label{current tensor}
J_{j}^{\vec{M}_{\xi}}=\frac{A}{\hbar M_{S}}\varepsilon_{\xi\mu\nu}\vec{M}_{\mu}\nabla_{j}\vec{M}_{\nu}.
\end{eqnarray}
 Latin indexes refer to the spatial components while Greek indices  to the spin projections. Since the magnetoelectric term is part of
 the effective field in Eq.(\ref{LLG}) it is expected to contribute to the spin current Eq.(\ref{current tensor}).
The temporal, spatially nonuniform temperature profile  $\nabla T\big(\vec{R},t\big)$ can be  inferred from the solution of the heat equation with the appropriate  source term  $I\big(\vec{R},t \big)$.  Explicitly this equation  reads  \cite{Etesami}:
\begin{eqnarray}
\label{heat equation}
\frac{\partial T\big(\vec{R},t\big)}{\partial t}=\frac{k_{ph}}{\rho C}\nabla^{2}T\big(\vec{R},t\big)+I\big(\vec{R},t \big).
\end{eqnarray}
 $C$ is the phonon heat capacity, $k_{ph}$ is the phononic thermal conductivity, and $\rho$ is the mass density.
The imparted energy, e.g. by laser pulses is usually not completely   absorbed by the system but is partially dissipated. Thermal loses can be
incorporated in a realistic modelling of the laser heating process. For more details we refer to Ref. \cite{Etesami}. It is important to consider the relevant  time scales.  When the characteristic time scale of the heating process is faster than the magnetization dynamics, (i.e. the phonon relaxation time scale and the time interval between laser pulses are shorter than the precession time $\tau<1/\gamma H_{eff}$) the magnetic system experiences an effective temperature which is deduced from an average over the much faster time scales.  In this case, instead of the coupled set of equations (\ref{thermal elastic wave}), (\ref{LLG}) and (\ref{heat equation}) we can explore a steady state problem.
After some algebra, we derive  for this case the solution of the elasticity equation for the displacement vector valid for an arbitrary averaged, non-uniform effective temperature as
\begin{eqnarray}
\label{solution}
\vec{u}\big(\vec{R}\big)=-\frac{\kappa \big(1+\sigma\big)}{12\pi \big(1-\sigma\big)}\nabla_{\vec{R}}\int\frac{T\big(\vec{R_{1}}\big)-T_{0}}{|\vec{R}-\vec{R_{1}}|}d^{3}\vec{R_{1}}.
\end{eqnarray}
The spatial temperature profile $T\big(\vec{R}\big)$ is arbitrary satisfying the  asymptotic boundary condition
$T_{0}=T\big(|\vec{R}-\vec{R}_0|\rightarrow\infty\big)$, where $\vec{R}_0$ defines the region where the heat source is localized.
In what follows we consider two different temperature profiles formed in the system due to the laser heating.

\subsection{Point-like heat sources}
Let us assume that the energy pumped for instance via a  laser irradiation  is localized such that $T\big(\vec{R}_{1}\big)-T_{0}=\frac{Q}{C}\delta\big(\vec{R}_{1}\big)$, where $Q$ is the heat  released by the laser, and $C$ is the heat capacity of the material. The displacement vector $\vec{u}\big(\vec{R}\big)$ reads for this case
\begin{eqnarray}
\label{1solution}
\vec{u}\big(\vec{R}\big)=\frac{\kappa\big(1+\sigma\big)Q}{12\pi \big(1-\sigma\big)C}\frac{\vec{R}}{R^{3}}.
\end{eqnarray}
With eq. (\ref{1solution})  we obtain  the explicit form of the deformation tensor
\begin{eqnarray}
\label{deformation tensor point}
\vec{\varepsilon}_{\xi \zeta}\big(\vec{R}\big)=\frac{\kappa\big(1+\sigma\big)}{12\pi\big(1-\sigma\big)}\frac{Q}{C}\frac{1}{R^3}
\bigg\{\delta_{\xi\zeta}-\frac{x_{\xi}x_{\zeta}}{R^{2}}\bigg\}.
\end{eqnarray}
 {Point heat sources} are an idealization.  In reality for   thin films the temperature profile decays exponentially, as proved by the exact numerical solution of Eq.(\ref{heat equation}).
Therefore,  we explore  an exponential temperature profile.
\subsection{Extended  heat sources}
Let us consider an exponential temperature profile matching the  numerical solutions of the heat equation with a non-local, i.e. extended  heating source $I$.
$T\big(\vec{R}_{1}\big)=\frac{Q_{1}}{C}\exp\big(-\beta|\vec{R}_{1}|\big)+T_{0}$ with the characteristic decay length $\beta$ and $Q_{1}$ is the density of the  heat released by the laser in the vicinity of the point $\vec{R}_{1}=0$. The temperature in the heating point $T(0)=Q_{1}/C+T_{0}$ and in the asymptotic $T\big(|\vec{R}_{1}|\rightarrow\infty\big)=T_{0}$.
Upon some  involved calculations for the displacement vector we infer
\begin{eqnarray}
\label{2solution}
\vec{u}\big(\vec{R}\big)=-\frac{\kappa\big(1+\sigma\big)Q_{1}}{3 \big(1-\sigma\big)C}\nabla_{\vec{R}}\nonumber\\
\bigg\{\frac{2}{\beta^{3}R}-e^{-\beta R}\bigg(\frac{2}{\beta^{3} R}+\frac{1}{\beta^2}\bigg)\bigg\}.
\end{eqnarray}
With this relation we obtain an explicit form of the deformation tensor as
\begin{eqnarray}
\label{deformation tensor explicit}
\displaystyle{\vec{\varepsilon}_{\xi \zeta}\big(\vec{R}\big)=-\frac{\kappa\big(1+\sigma\big)Q_1}{3\big(1-\sigma\big)C}}\nonumber \\
\bigg\{\delta_{\xi\zeta}F_{1}\big(\beta R\big)
+\frac{x_{\xi}x_{\zeta}}{R^{2}}F_{2}\big(\beta R\big)\bigg\}.
\end{eqnarray}
For brevity  we introduced the notations $$F_{1}(y)=\big[\exp(-y)\big(1+2/y\big)-2/y^{2}\big(1-\exp(-y)\big)\big]/y$$ and
$$F_{2}(y)=\big[6-6\exp(-y)-y\exp(-y)\big(6+3y+y^{2}\big)\big]/y^{3}.$$\\
Formally Eqs. (\ref{1solution})-(\ref{deformation tensor explicit}) exhibit  a  nonphysical divergence in the limit $\vec{R}\rightarrow 0$.  We note however, in a coarse-grained approach the unit cell is non-deformable. Therefore, the  minimal $\vec{R}\rightarrow 0$ for which the study of the deformation makes sense is larger than the size of the coarse-grained  cell $\mid\vec{R}\rightarrow 0\mid>a$ and $a\approx 10$ nm.
\begin{figure}
	\centering
	\includegraphics[width=0.49\textwidth]{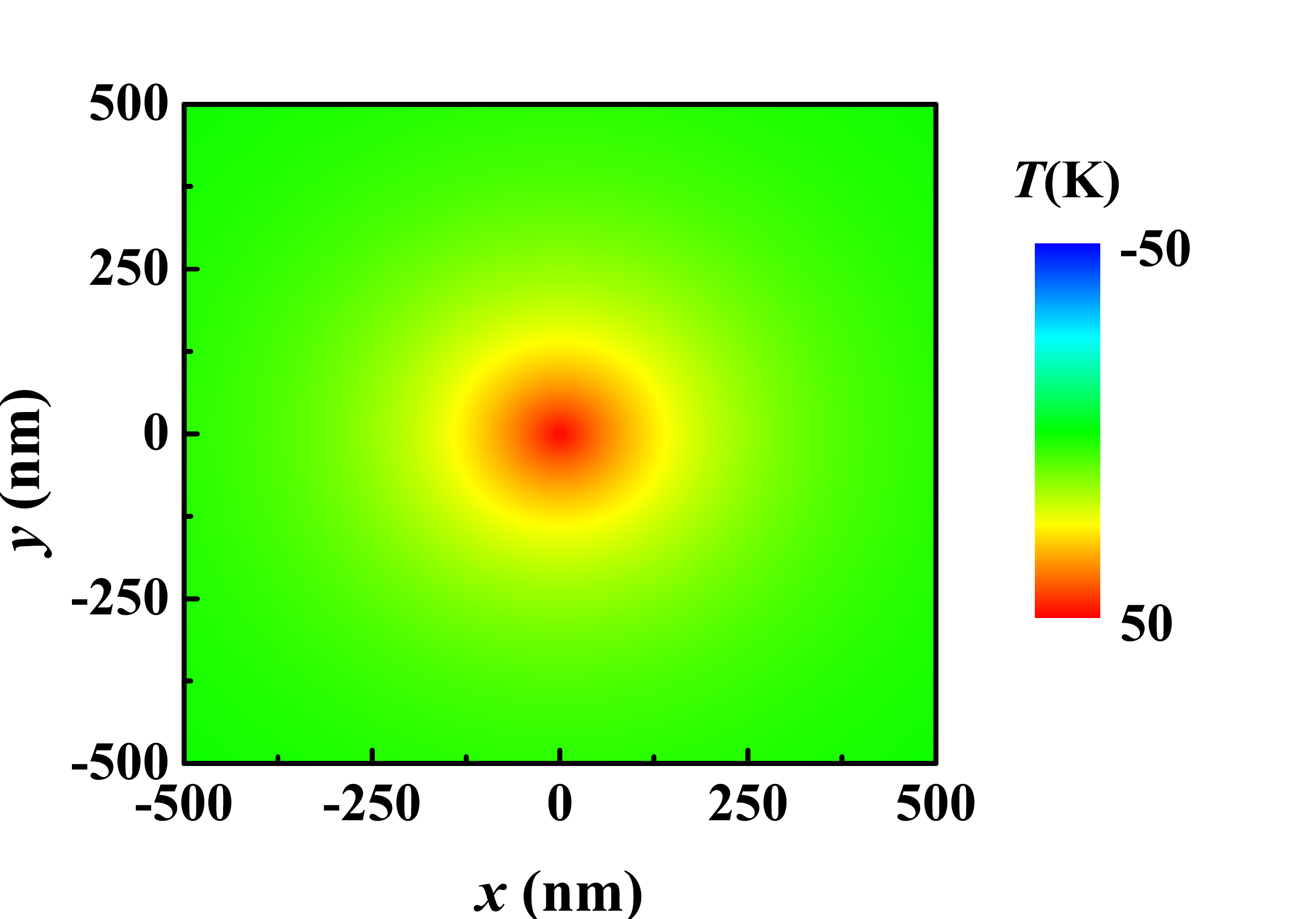}		
	\caption{The temperature profile in the 3D bulk system $T\big(\vec{R_{1}\big)}=\frac{Q_{1}}{C}\exp(-\beta\mid\vec{R_{1}}\mid)$, formed due to the effect of extended heat source.
The density of the heat released by the laser in the vicinity of the point $R_{1} = 0$ is equal to $Q_{1} / C= 50$ K, the heat capacity of the material Nickel $C=502$ J/(kg K). For convenience we present projection of the temperature profile on the $XOY$ plane by setting $z=0$.}
	\label{Tem-num}
\end{figure}

 \begin{figure}
	\centering
	\includegraphics[width=0.49\textwidth]{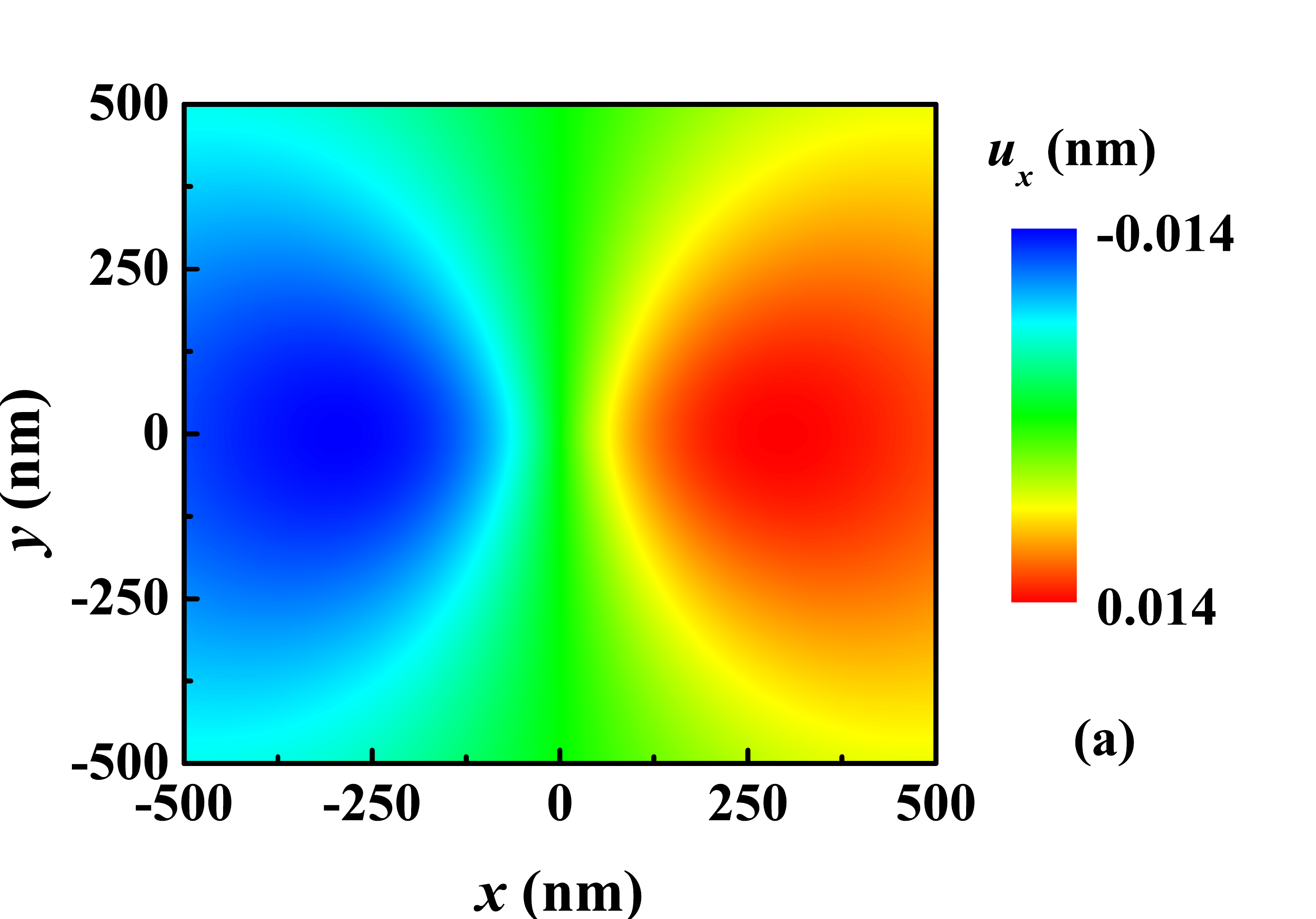}	
	\includegraphics[width=0.49\textwidth]{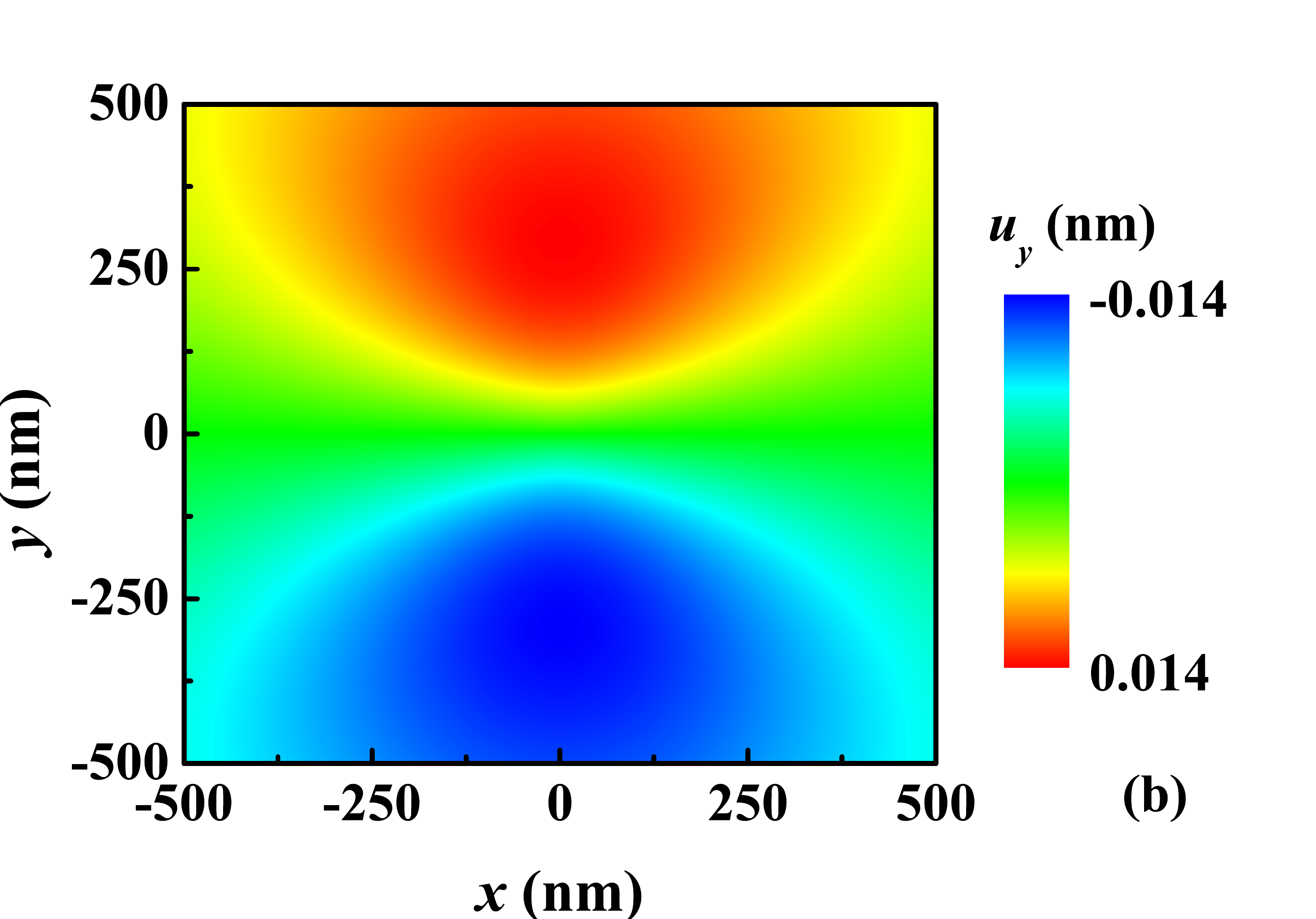}		
	\caption{Displacement tensors $ u_x $ (a) and $ u_y $ (b) in the 3D bulk system, induced by the exponential temperature profile.  The symmetry properties of the displacement tensors $ u_x $ and $ u_y $.  In particular, the $u_{x}$ component manifests  a mirror symmetry with respect to the reflection $y\rightarrow - y$ and antisymmetry with respect to the reflection $x\rightarrow - x$. Concerning the component $ u_y $ the behavior  is opposite: we find symmetry with respect to the reflection $x\rightarrow - x$ and antisymmetry with respect to the reflection $y\rightarrow - y$.}
	\label{disu-num}
\end{figure}
While in general case expression for displacement vector Eq.(\ref{2solution}) is quite involved, easy to see that
in the asymptotic limit of the large $R\mapsto\infty$ we have a decay $1/R^{2}$.

\begin{figure}
	\centering
	\includegraphics[width=0.49\textwidth]{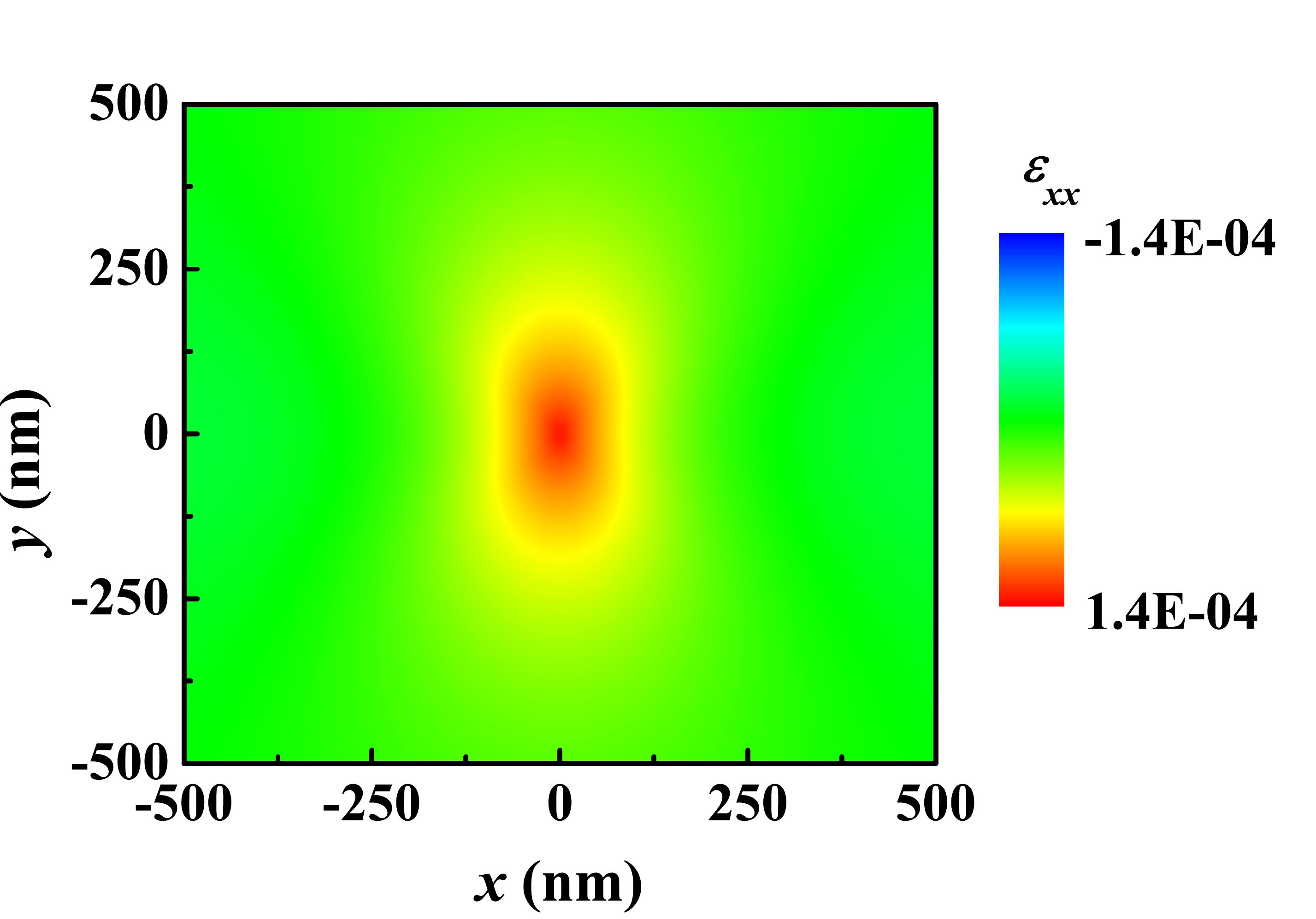}			
	\caption{The deformation tensor in the 3D bulk system $ \epsilon_{xx} $ induced by an exponential temperature profile (extended heat source) $T\big(\vec{R_{1}\big)}=\frac{Q_{1}}{C}\exp(-\beta\mid\vec{R_{1}}\mid)$. The density of the heat released by the laser $Q_{1} / C= 50$ K, the heat capacity of the material Nickel $C=502$ J/(kg K).}
	\label{eps-numxx}
\end{figure}

\subsection{Linear Temperature profile}

Linear temperature profile has particular interest for spin Seebeck experiments \cite{Uchidanew, Adachinew, Xiaonew}.
We consider linear temperature profile of the following form $T(R)=-\big(T'-T_{0}\big)R/R_{max}+T'$, where $T'>T_{0}$, and the temperature at the edges is equal to $T(0)=T'$,
$T(R_{max})=T_{0}$. After implementing liner temperature profile, for displacement vector we deduce:
\begin{eqnarray}
\label{DeformationLinearProfile}
\vec{u}(\vec{R})=\frac{\kappa (1+\sigma)(T'-T_{0})\vec{R}}{9(1-\sigma)}\bigg(1-\frac{3R}{4R_{max}}\bigg),
\end{eqnarray}
and for the deformation tensor we have:
\begin{equation}
\begin{split}
\label{deformationtensorlinearprofile}
& \varepsilon_{\xi\zeta}=\frac{\kappa(1+\sigma)(T'-T_{0})}{9(1-\sigma)}\times\\
& \bigg(\delta_{\xi\zeta}\big(1-3R/4R_{max}\big)-\frac{3x_{\xi}x_{\zeta}}{4R_{max}R}\bigg).
\end{split}
\end{equation}
As we see in case of linear temperature gradient asymptotic behavior of the displacement vector and deformation tensor is different
and non-monotonic in $R=\sqrt{x^{2}+y^{2}+z^{2}}$. Maximum of the absolute value of the displacement vector corresponds to the case
$|\vec{u}(\vec{R})|=2R_{max}/3$.

\section{Dispersion relations for thermal magneto-elastic spin waves in  bulk systems}

Taking into account  Eq.(\ref{LLG}) and Eq.(\ref{1solution}) - Eq.(\ref{deformation tensor explicit}) and assuming that the ground state magnetization is aligned parallel to the $z$ axis we derive the following dispersion relation for the coupled  magnetoelastic magnonic modes in the 3D bulk system

\begin{equation}
\label{magnon phonon modes01}
\begin{split}
\displaystyle\omega^{2}\big(q,\vec{R}\big)=
\\& \bigg(\frac{2\gamma A_{ex}}{M_{s}^2}q^{2}+K+\gamma H_{0}-\frac{2\gamma B_{12}}{M_{s}^2}\big(\varepsilon_{zz}-\varepsilon_{xx}\big)\bigg)\times
\\& \bigg(\frac{2\gamma A_{ex}}{M_{s}^2}q^{2}+K+\gamma H_{0}-\frac{2\gamma B_{12}}{M_{s}^2}\big(\varepsilon_{zz}-\varepsilon_{yy}\big)\bigg)-
\\& \bigg(\frac{2\gamma B_{12}}{M_{s}^2}\varepsilon_{xy}\bigg)^{2}.
\end{split}
\end{equation}
Here $B_{1}=B_{2}=B_{12}$  are the magnetoelastic coupling constants.

Obviously  in the absence of the magnetoelastic effect $\vec{\varepsilon}_{\xi \zeta}\big(\vec{R}\big)=0$, the obtained result
falls back to the well-known magnonic dispersion relation. Depending on the values of the components of deformation tensor $\vec{\varepsilon}_{\xi \zeta}\big(\vec{R}\big)$ the
magnetoelastic contribution in the magnonic dispersion relations can be positive or negative. A negative contribution $-\frac{2\gamma B_{12}}{\mu_{0}M_{s}}\big(\varepsilon_{zz}-\varepsilon_{xx}\big)<0$
and $-\frac{2\gamma B_{12}}{\mu_{0}M_{s}}\big(\varepsilon_{zz}-\varepsilon_{yy}\big)<0$  decreases the magnonic gap, while a
positive contribution leads to an enhancement. Thus, the thermal magnetoelastic effect can be used as  a tool for reducing the magnonic gap imposed by the magnetocrystalline anisotropy or by an external magnetic field.
A reduction of the gap naturally increases the spin Seebeck effect since it enhances the number of magnons contributing to the spin current. We note that $\vec{\varepsilon}_{\xi \zeta}\big(\vec{R}\big)$
is a local quantity and can be different for different $\vec{R}$. In some particular cases  the magnonic gap can be enhanced and this naturally decreases the spin Seebeck current.
After inserting the explicit expression for the deformation tensor $\vec{\varepsilon}_{\xi \zeta}\big(\vec{R}\big)$ in
Eq. (\ref{magnon phonon modes01}) we deduce

\begin{equation}
\label{magnon phonon modes}
\displaystyle\omega\big(q,\vec{R}\big)=
\frac{1}{\hbar}\sqrt{\big(Aq^{2}+ f^{(1)}_{R}\big)\big(Aq^{2}+f^{(2)}_{R}\big)-f^{(3)}_{R}}.
\end{equation}
The  particular values of the  introduced functions  $f^{(1,2,3)}_{R}$ are different for the exponential, the point-like heat source, or for the linear temperature profile.\\

a) In case of a point-like heat source we find:
\begin{equation}
\label{fp1}
f_{R}^{(1)}=\gamma\hbar\bigg\{\frac{\kappa(1+\sigma)}{6\pi M_{s}^2(1-\sigma)}\frac{1}{R^2}\frac{Q}{C}B_{1}(z^2-x^2)+H_{0}\bigg\},
\end{equation}
\begin{equation}
\label{fp2}
f_{R}^{(2)}=\gamma\hbar\bigg\{\frac{\kappa(1+\sigma)}{6\pi M_{s}^2(1-\sigma)}\frac{1}{R^2}\frac{Q}{C}B_{1}(z^2-y^2)+H_{0}\bigg\},
\end{equation}
\begin{equation}
\label{fp3}
f_{R}^{(3)}=\bigg\{\gamma\hbar\frac{\kappa(1+\sigma)}{6\pi M_{s}^2(1-\sigma)}\frac{1}{R^2}\frac{Q}{C}B_{2}xy\bigg\}^2.
\end{equation}

b) In the case of an exponential temperature profile one finds:

\begin{equation}
\label{f1}
\begin{split}
\displaystyle f^{(1)}_{R}=\gamma \hbar\bigg\{\frac{2\kappa\big(1+\sigma\big)Q_{1}\big(z^{2}-x^{2}\big)}{3M_{s}^2R^{2}\big(1-\sigma\big)C}B_{1}F_{2}\big(\beta R\big)+H_{0}\bigg\},
\end{split}
\end{equation}

\begin{equation}
\label{f2}
\begin{split}
\displaystyle f^{(2)}_{R}=\gamma \hbar\bigg\{\frac{2\kappa\big(1+\sigma\big)Q_{1}\big(z^{2}-y^{2}\big)}{3M_{s}^2R^{2}\big(1-\sigma\big)C}B_{1}F_{2}\big(\beta R\big)+H_{0}\bigg\},
\end{split}
\end{equation}

\begin{equation}
\label{f3}
\begin{split}
\displaystyle f^{(3)}_{R}=\bigg\{\gamma \hbar \frac{2\kappa\big(1+\sigma\big)Q_{1} xy}{3M_{s}^2R^{2}\big(1-\sigma\big)C}B_{2}F_{2}\big(\beta R\big)\bigg\}^{2}.
\end{split}
\end{equation}

c) In the case linear temperature profile we deduce:

\begin{equation}
\label{fl1}
f_{R}^{(1)}=\hbar\gamma\bigg\{\frac{B_1}{6R_{max}RM_{s}^{2}}
\frac{\kappa(1+\sigma)(T_{0}-T')}{(1-\sigma)}(x^2-z^2)+H_{0}\bigg\},
\end{equation}
\begin{equation}
\label{fl2}
f_{R}^{(2)}=\hbar\gamma\bigg\{\frac{B_1}{6R_{max}RM_{s}^{2}}
\frac{\kappa(1+\sigma)(T_{0}-T')}{(1-\sigma)}(y^2-z^2)+H_{0}\bigg\},
\end{equation}
\begin{equation}
\label{fl3}
f_{R}^{(3)}=\bigg\{\hbar\gamma\frac{B_2}{6R_{max}RM_{s}^{2}}
\frac{\kappa(1+\sigma)(T_{0}-T')}{(1-\sigma)}xy\bigg\}^2.
\end{equation}
As we see from Eq.(\ref{magnon phonon modes}) - Eq.(\ref{fl3})  the dispersion relation for mixed magnon-phonon modes are rather complex.
The dependence on the  spatial variable $\vec{R}$ is non-uniform with an anisotropic character of the magnonic modes.
Note that obtained analytical results correspond to the 3D model, while for the sake of simplicity in numerical calculations we consider 2D model. \\
\begin{figure}
	\centering
	\includegraphics[width=0.49\textwidth]{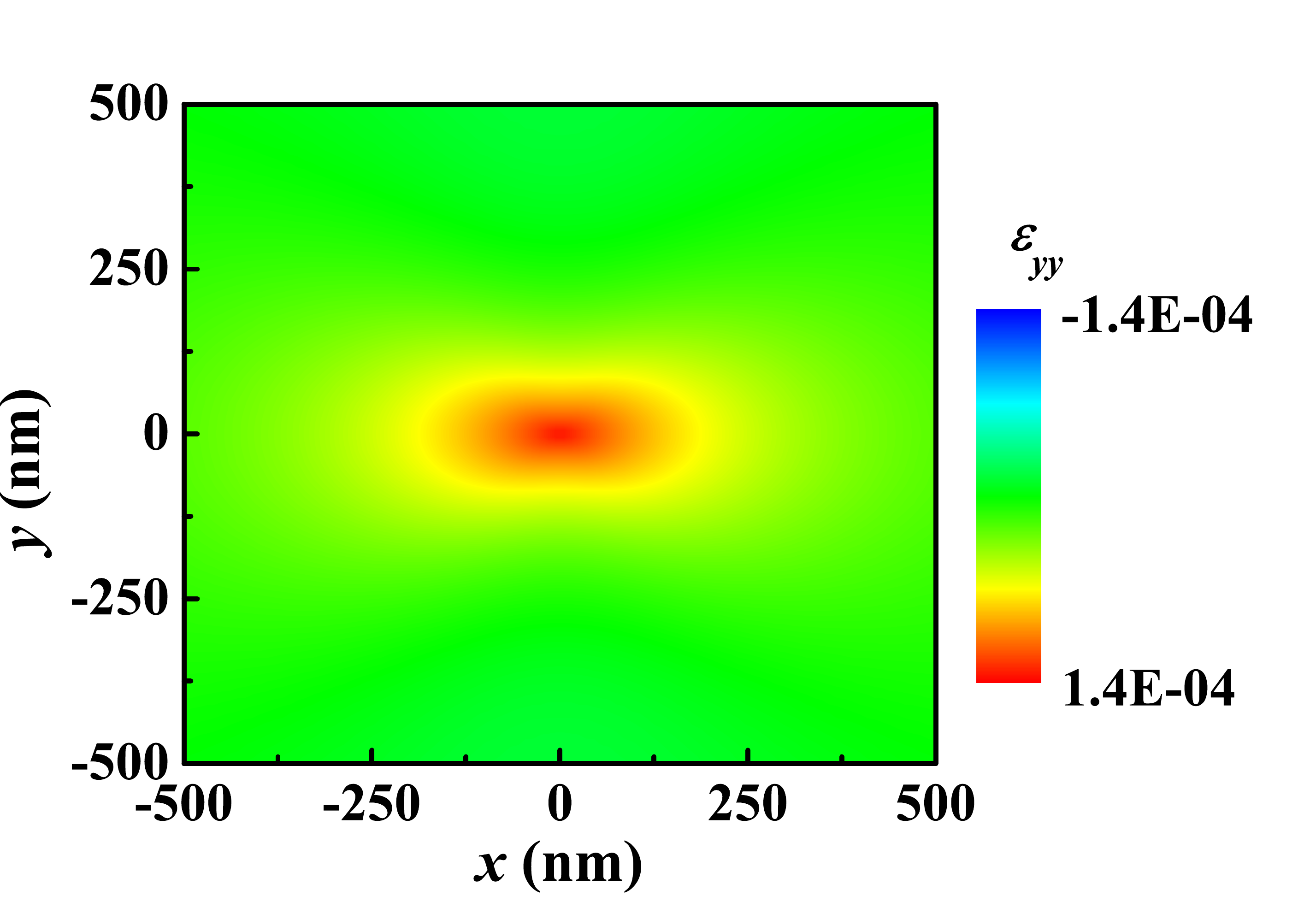}
	\caption{The deformation tensor in the 3D bulk system $ \epsilon_{yy} $, induced by an exponential temperature profile (extended heat source) $T\big(\vec{R_{1}\big)}=\frac{Q_{1}}{C}\exp(-\beta\mid\vec{R_{1}}\mid)$. The density of the heat released by the laser $Q_{1} / C= 50$ K, the heat capacity of the material Nickel $C=502$ J/(kg K). }
	\label{eps-numyy}
\end{figure}

\begin{figure}
	\centering
	\includegraphics[width=0.49\textwidth]{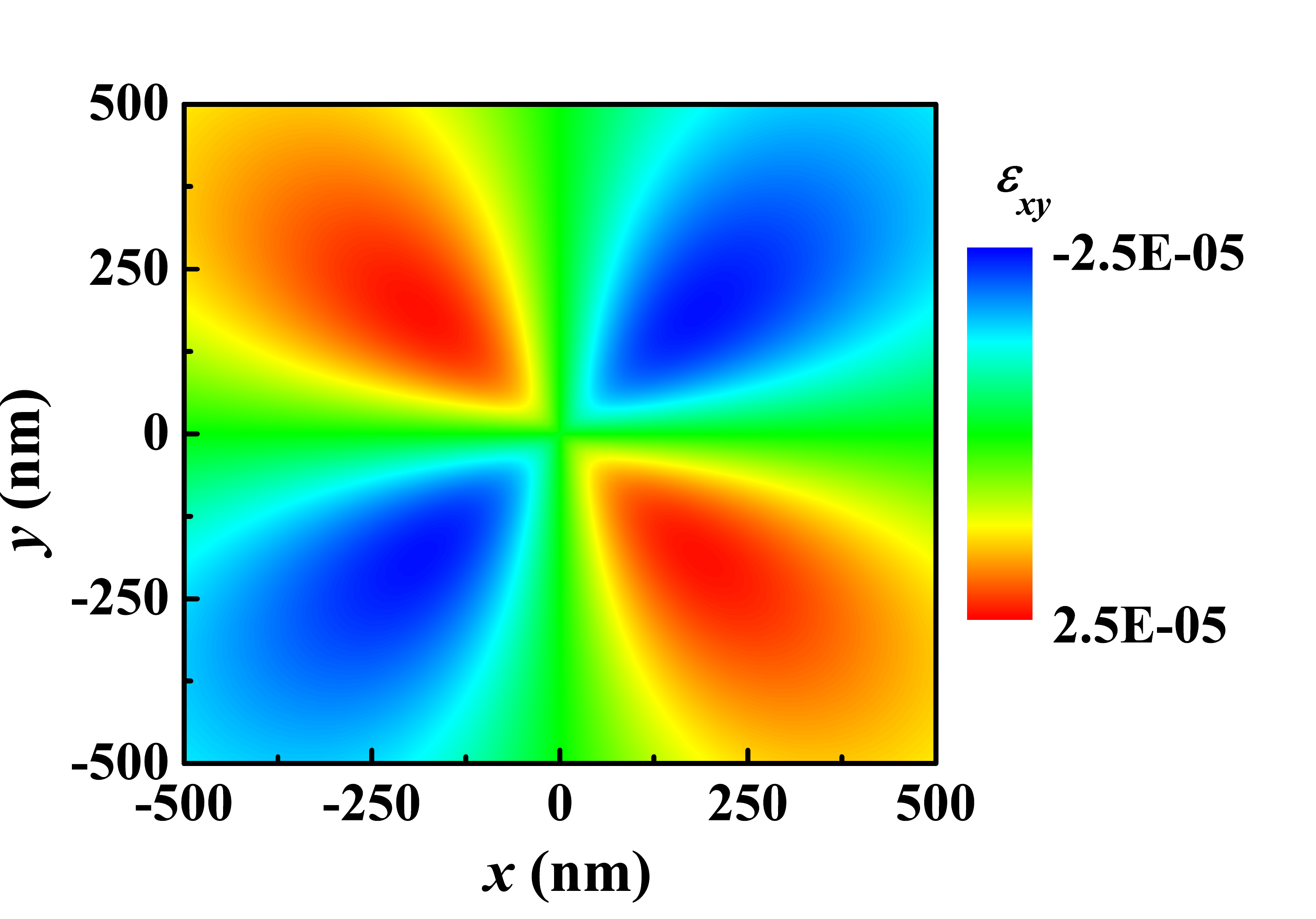}		
	\caption{The deformation tensor in the 3D bulk system $ \epsilon_{xy} $, induced by an exponential temperature profile (extended heat source) $T\big(\vec{R_{1}\big)}=\frac{Q_{1}}{C}\exp(-\beta\mid\vec{R_{1}}\mid)$. The density of the heat released by the laser $Q_{1} / C= 50$ K, the heat capacity of the material Nickel $C=502$ J/(kg K). }
	\label{eps-numxy}
\end{figure}

\begin{figure}
	\centering
	\includegraphics[width=0.49\textwidth]{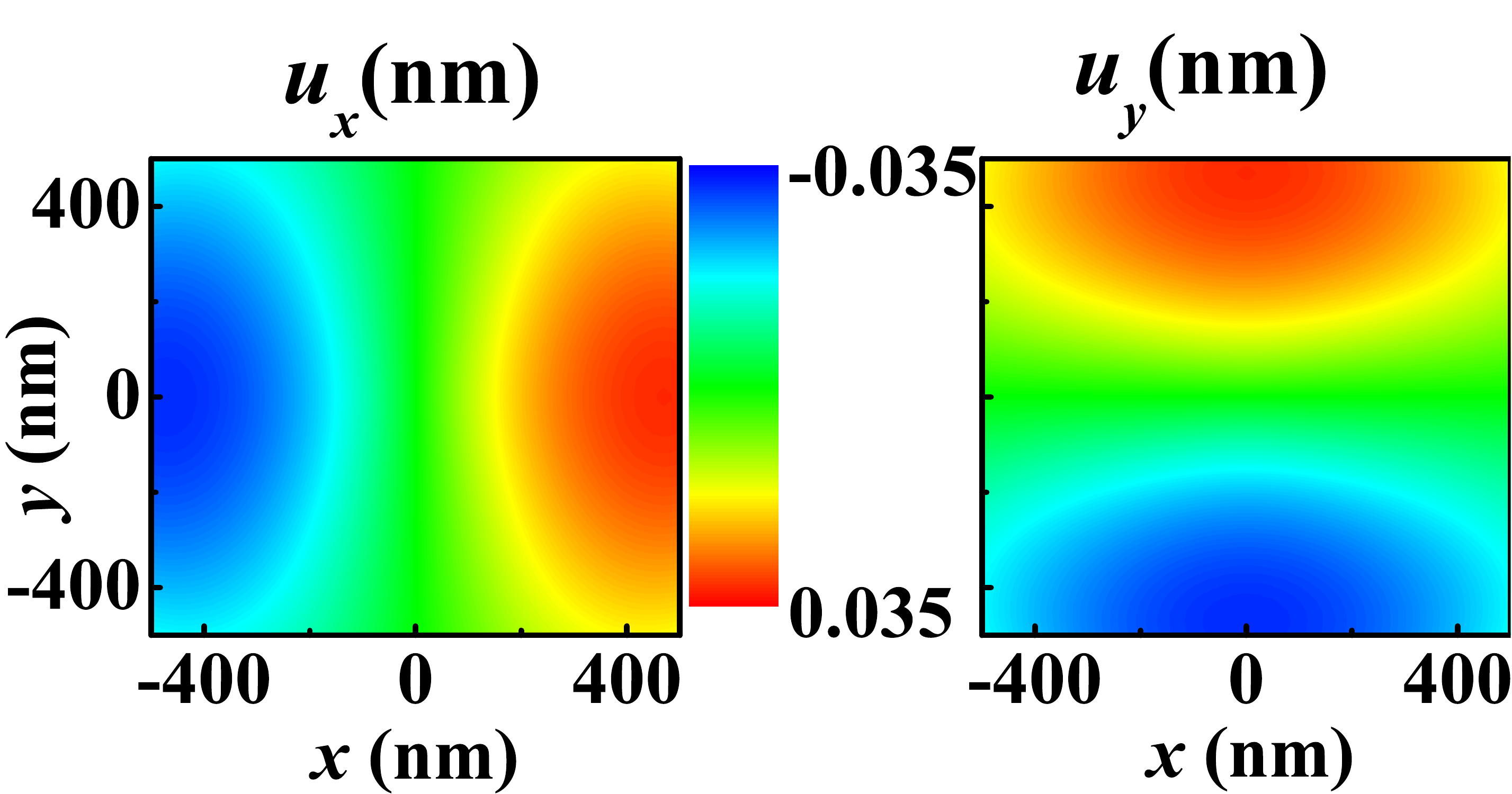}		
	\caption{The displacement vector in the 3D bulk system $ u_x $ and $ u_y $, induced by a linear temperature profile. The temperature in the center is equal to $ T(0) = 50 $ K and at the edges $ T(R_{max}) = 0 $.}
	\label{3Dlinearnumxy}
\end{figure}

\begin{figure}
	\centering
	\includegraphics[width=0.49\textwidth]{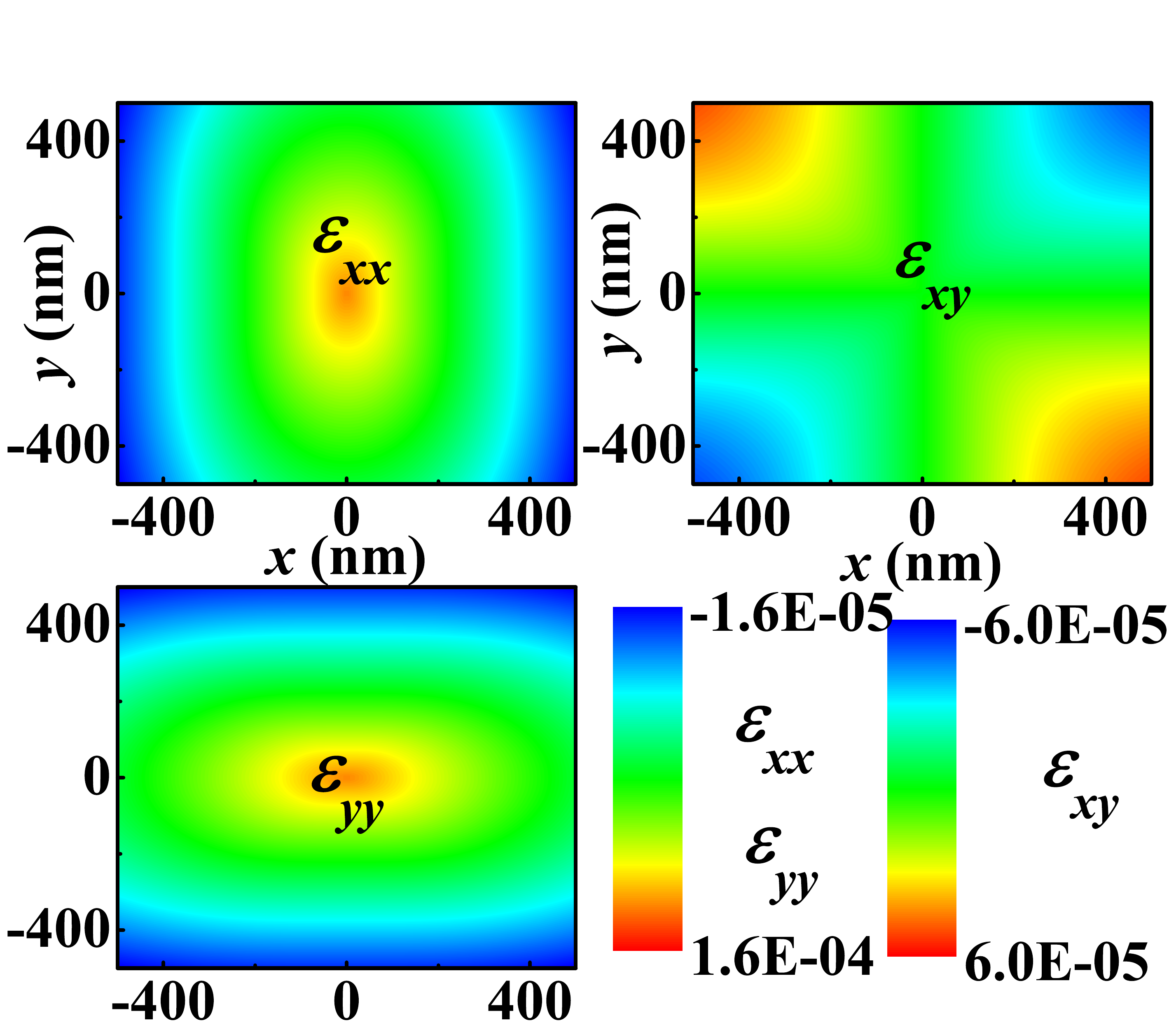}		
	\caption{The deformation tensors in the 3D bulk system $ \epsilon_{xx} $, $ \epsilon_{yy} $ and $ \epsilon_{xy} $, induced by a linear temperature profile. The temperature in the center is equal to $ T(0) = 50 $ K and at the edges $ T(R_{max}) = 0 $.}
	\label{3Dlineartensor}
\end{figure}

  \begin{figure}
	\centering
	\includegraphics[width=0.49\textwidth]{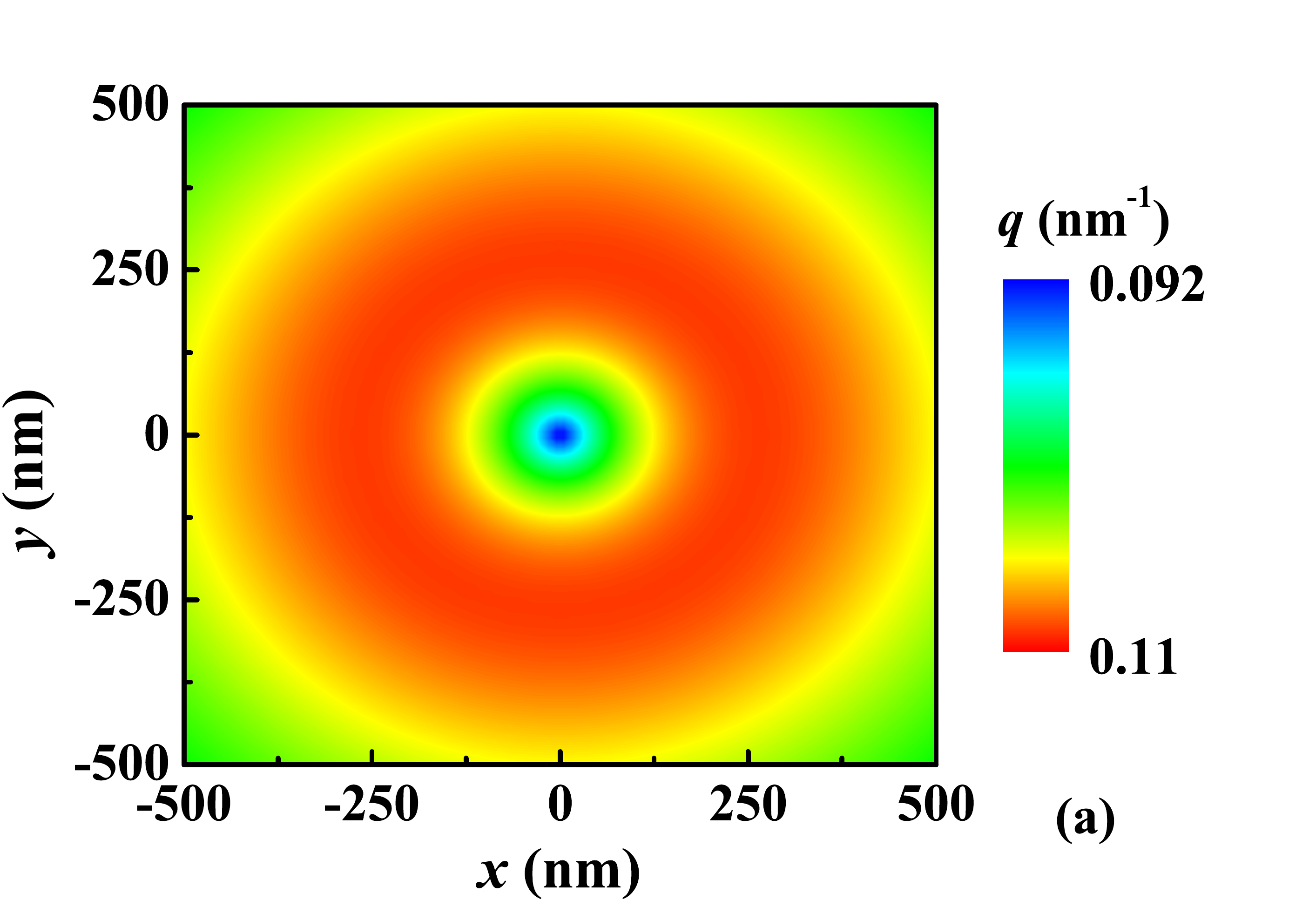}	
	\includegraphics[width=0.49\textwidth]{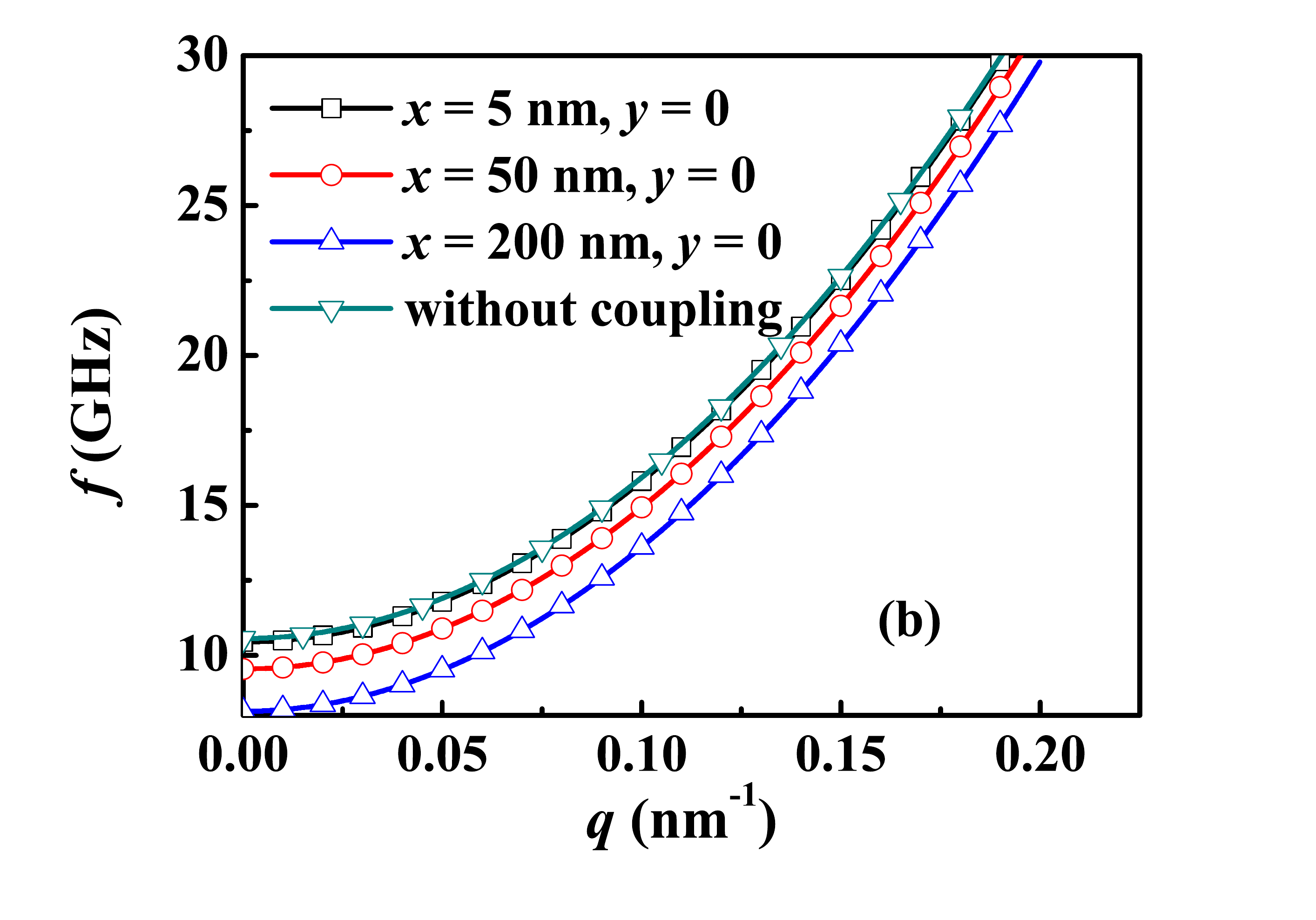}		
	\caption{ (a) The profile of the wave-vector $ q(x,y) $ in the 3D bulk system $ f = \omega\big(q,\vec{R}\big)/ (2\pi)$. Spatial distribution of magnons with the same fixed frequency $f = \omega\big(q,\vec{R}\big)/ (2\pi)= 15$  GHz but different wave vectors $q$ is plotted. (b) For selectively chosen areas: $ x = 5, 50, 200 $ nm, $z = 0$  and $ y = 0 $, the dispersion relation of magnons is calculated using analytical result Eq. (\ref{magnon phonon modes}). Values of the parameters $A_{ex} = 4.6 \times 10^{-12}$ J/m, $\gamma=1.76 \times 10^{11}$ 1/(T s), $M_{s}=4.8 \times 10^{5}$ A/m, $H_{0}= 3 \times 10{5}$ A/m, $B_{1} = B_{2} = 7.85 \times 10^{8}$ J/m$ ^3 $.}
	\label{dispersion relation}
\end{figure}

\begin{figure}
	\centering
	\includegraphics[width=0.49\textwidth]{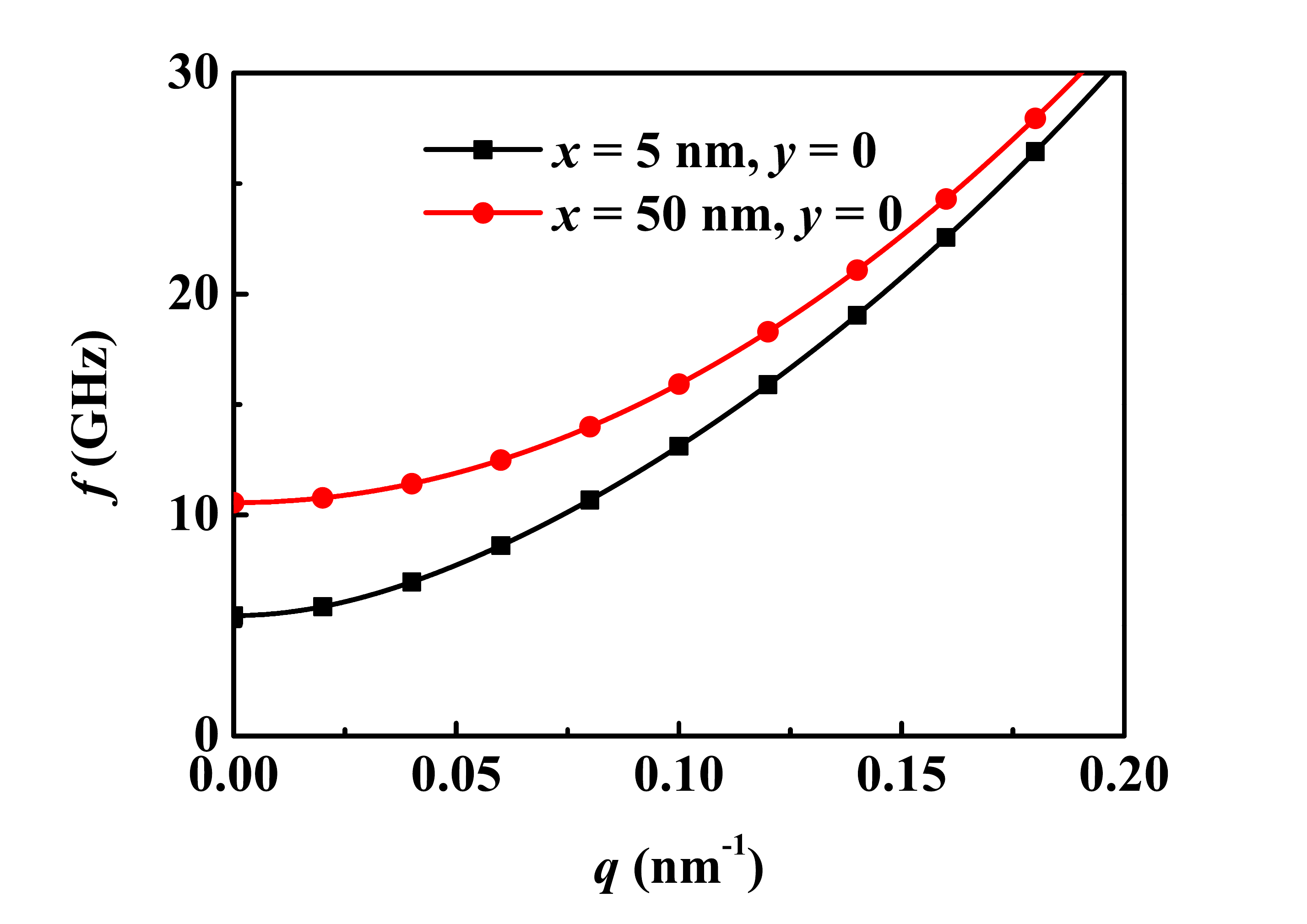}	
	\caption{2D thin magnetic film. For $ x = 5, 50 $ nm and $ y = 0 $, the dispersion relation calculated using Eq. (\ref{Dispersion2D})
and the the following values of parameters $A_{ex} = 4.6 \times 10^{-12}$ J/m, $\gamma=1.76 \times 10^{11}$ 1/(T s), $M_{s}=4.8 \times 10^{5}$ A/m, $H_{0}= 3 \times 10{5}$ A/m, $B_{1} = B_{2} = 7.85 \times 10^{8}$ J/m$ ^3 $, {$ -I_0 / \alpha = 0.01 $ K}, $ \alpha = 5 \times 10^6 $ m$ ^{-1} $, $ \kappa = 1.3 \times 10^{-5} $ K$ ^{-1} $ and $ \sigma = 0.31 $.}
	\label{fig_3-2d}
\end{figure}

In prior to the numerical calculations, we present illustrations to support involved analytical findings.
We adopted the material parameters of Nickel: the saturation magnetization is $ M_s = 4.8 \times 10^5 $ A/m, the exchange constant is $ A_{ex} = 4.6 \times 10^{-12} $ A/m, the damping constant is $ \alpha =0.01 $,  the mass density is $ \rho = 8908 $ kg/m$ ^3 $, the heat capacity is $ C =502  $ J/(kg K), the thermal conductivity is $ k_{ph}= 91 $ W/(m K),   Young's modulus is $ E = 200 $ GPa, the Poisson's ratio is $ \sigma = 0.31 $, and the linear thermal expansion coefficient is $ \kappa = 1.3 \times 10^{-5} $ K $^{-1}$. For an exponential temperature profile we set the decay length  as $ \beta = 5 \times 10^6 $m$ ^{-1} $, $ T_0 = 0 $ and $Q_{1}/C = 50 $ K.
The result for the exponential temperature profile $T\big(\vec{R}_{1}\big)=\frac{q}{C}\exp\big(-\beta|\vec{R}_{1}|\big)+T_{0}$ is shown in Fig. \ref{Tem-num}. As we see the temperature profile is isotropic and symmetric in the $xy$ plane. The temperature is maximal in the area heated by laser and  decays exponentially with increasing distance from the laser spot.
The symmetry properties of the displacement tensors $ u_x $ and $ u_y $ for an exponential temperature profile are quite  intriguing, see Fig. \ref{disu-num}.
We clearly observe  that the $u_{x}$ component possesses a mirror symmetry with respect to  reflection $y\rightarrow - y$, and is antisymmetric with respect to the reflection $x\rightarrow - x$. Concerning the component $ u_y $, the situation is opposite: it is  symmetric with respect to the reflection $x\rightarrow - x$ and antisymmetric with respect to $y\rightarrow - y$.
In case of the linear temperature gradient see Fig.\ref{3Dlinearnumxy} symmetry properties of the displacement vector are preserved, but maximum is slightly shifted to the edges of the sample
$|\vec{u}(\vec{R})|=2R_{max}/3$. The components of the deformation tensor $ \epsilon_{xx} $, $ \epsilon_{yy} $ and $ \epsilon_{xy} $ for the extended heat source are shown in Fig. \ref{eps-numxx}, Fig. \ref{eps-numyy}, Fig. \ref{eps-numxy} and for the linear temperature profile in Fig. \ref{3Dlineartensor} . The diagonal components $ \epsilon_{xx} $ and $ \epsilon_{yy} $ are larger but localized, while the non-diagonal component of the deformation tensor $ \epsilon_{xy} $ decays slower with distance  and is finite in the whole sample.

\begin{figure}
	\centering
	\includegraphics[width=0.49\textwidth]{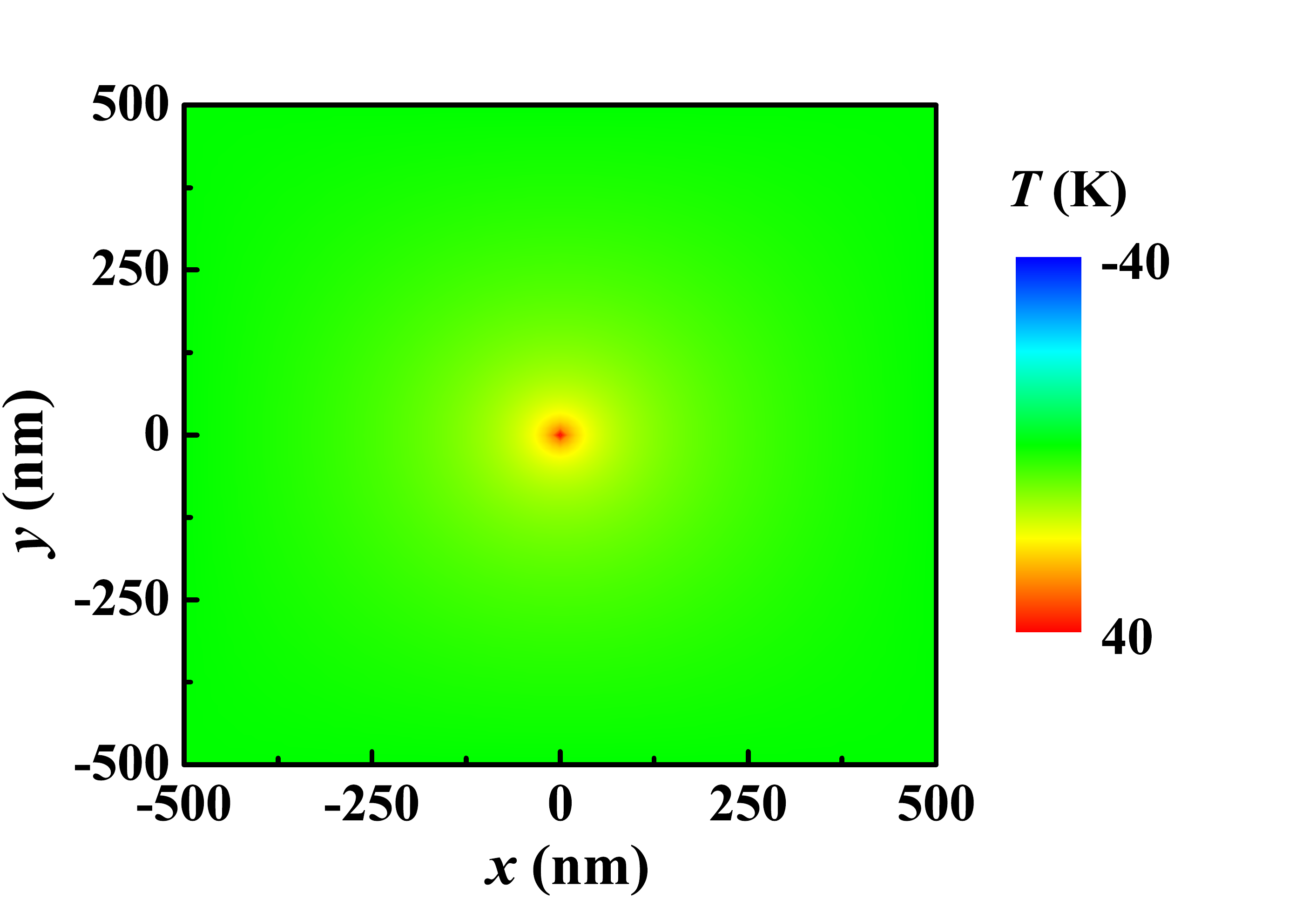}		
	\caption{2D thin magnetic film. The steady state spatial temperature profile formed in the system due to a laser heating.
		The result is obtained via a numerical solution of the heat equation.  The maximum temperature in the area heated by laser pulses is
		$ T_{0} = 40 $K in the vicinity of the point $ x = 0 $ and $ y = 0 $ and  decays gradually to zero with the distance. }
	\label{fig_1}
\end{figure}

\begin{figure}
	\centering
	\includegraphics[width=0.49\textwidth]{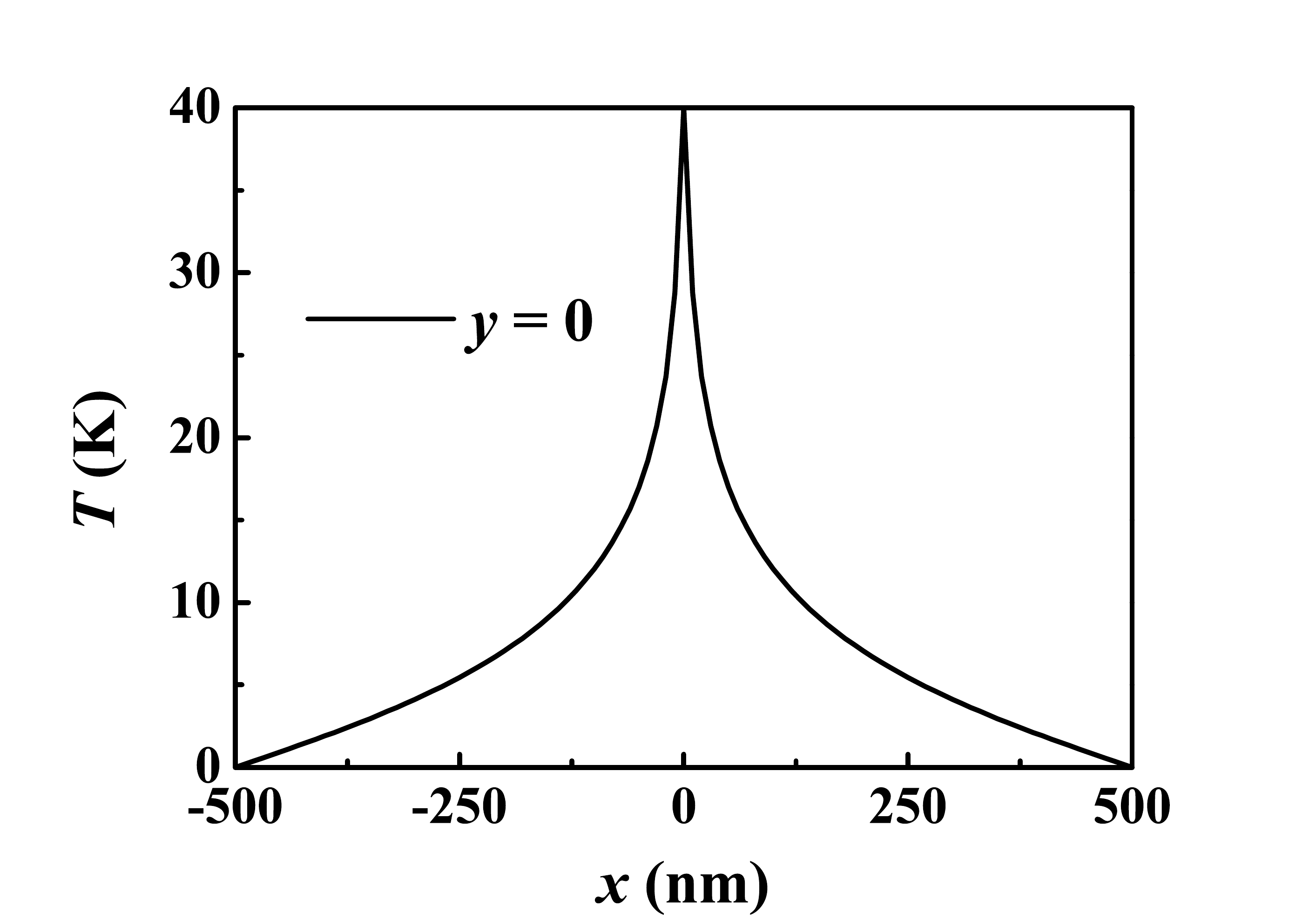}		
	\caption{2D thin magnetic film. Projective plot of the steady state spatial temperature profile formed in the system due to  laser heating as follows from the numerical solution of the heat equation.}
	\label{fig_2}
\end{figure}

\begin{figure}
	\centering
	\includegraphics[width=0.49\textwidth]{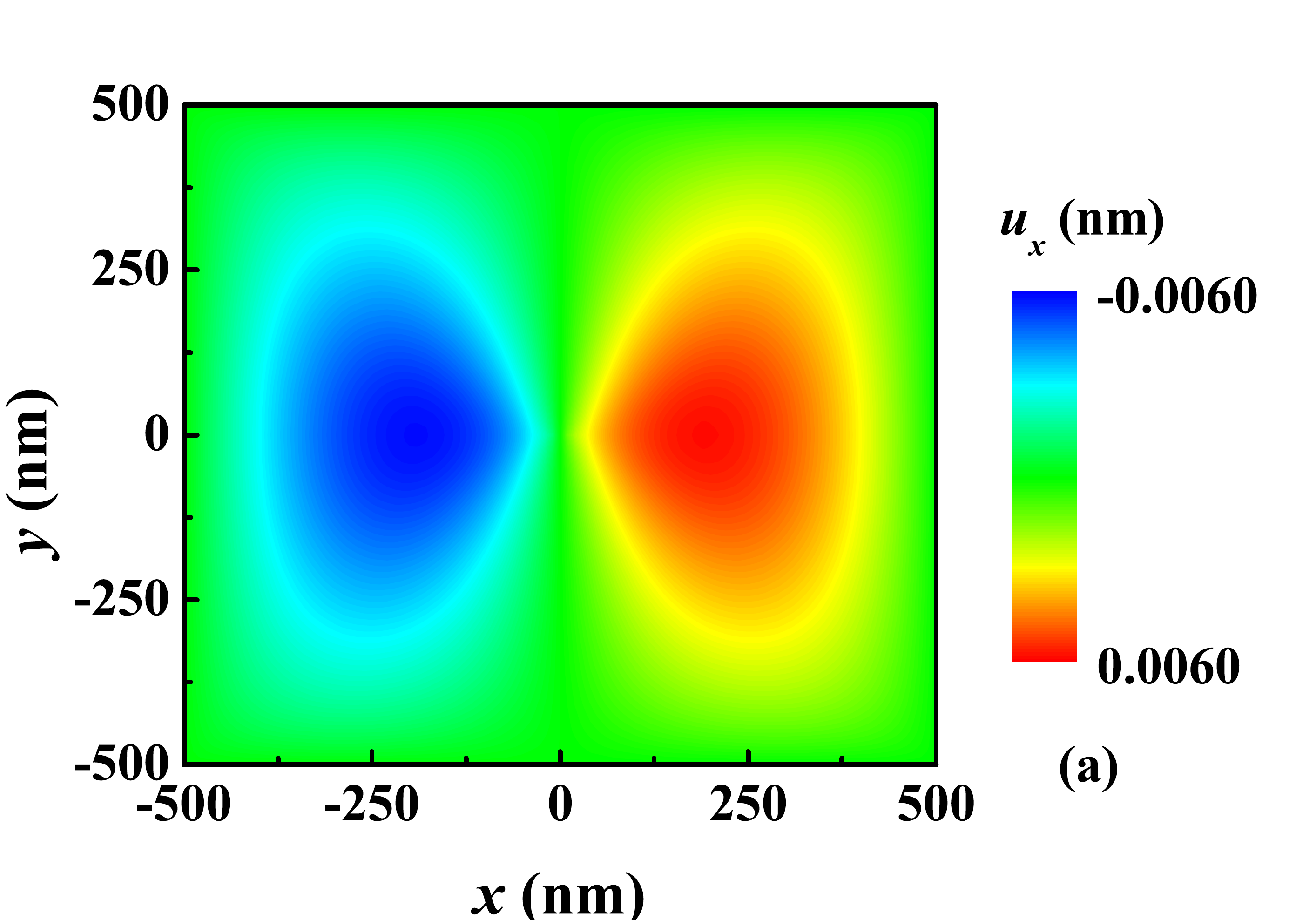}	
	\includegraphics[width=0.49\textwidth]{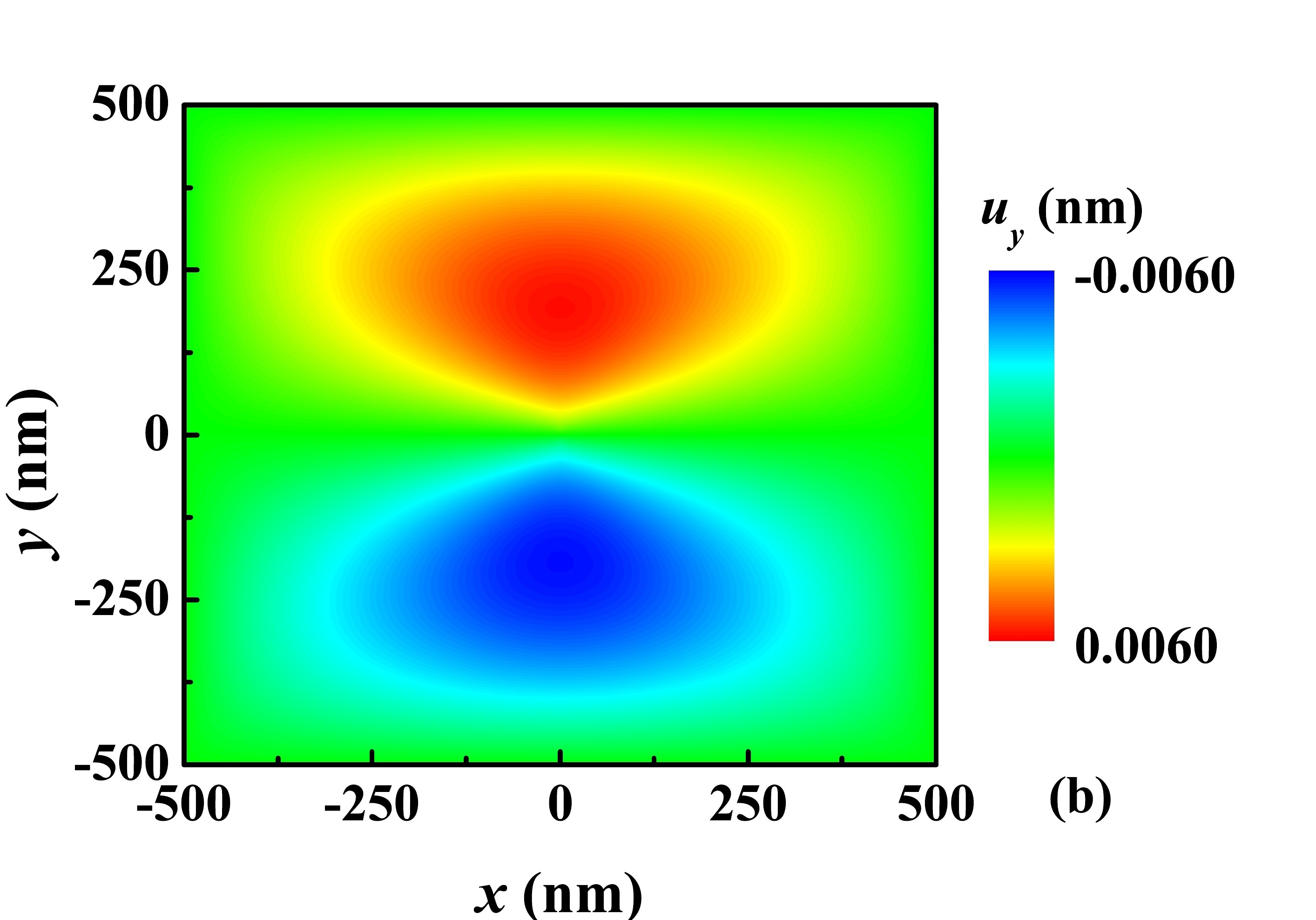}	
	\caption{2D thin magnetic film. Steady state spatial profile of the displacement vector (a) $ u_x $ and (b) $ u_y $. }
	\label{fig_3}
\end{figure}

\begin{figure}
	\centering
	\includegraphics[width=0.49\textwidth]{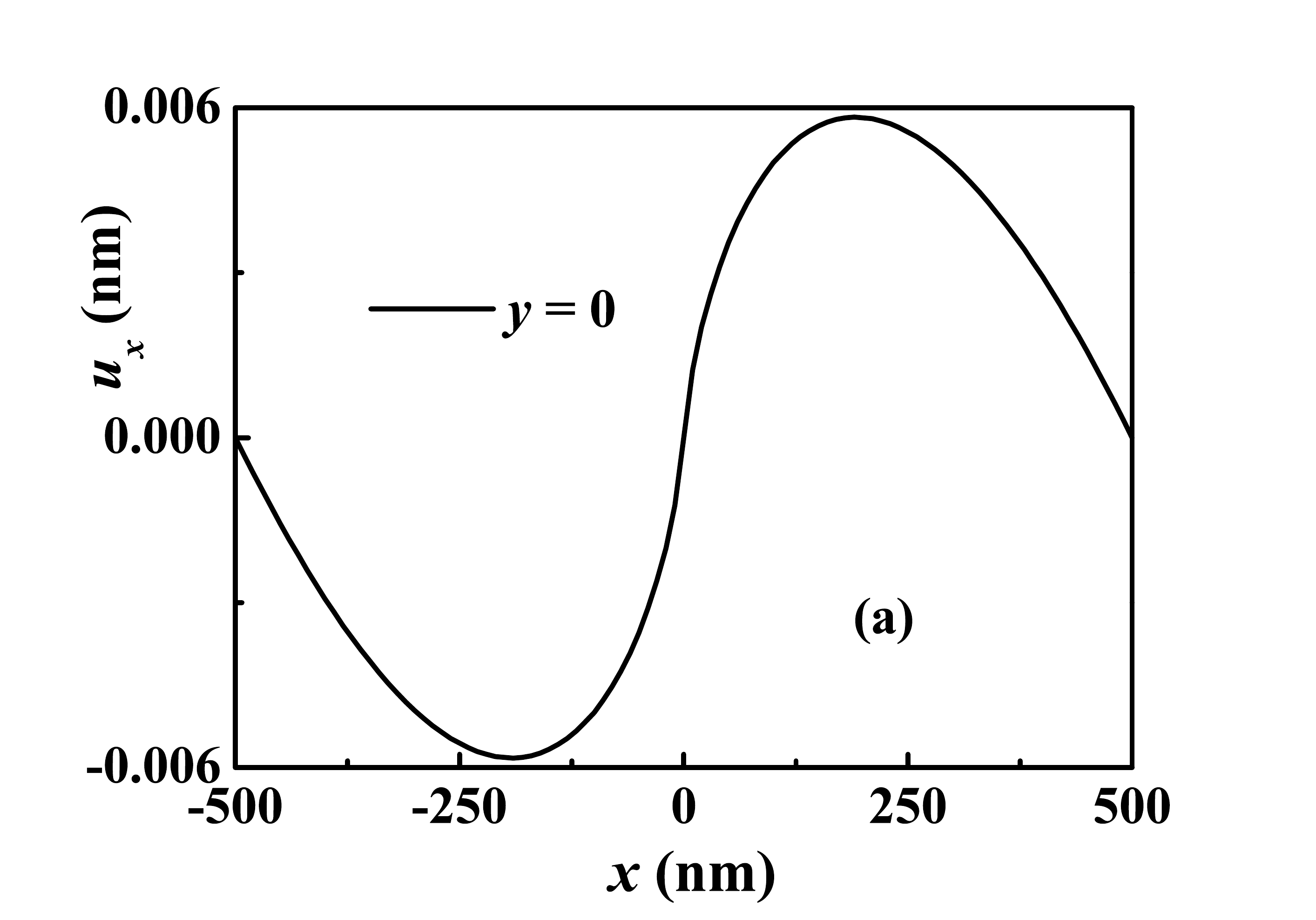}	
	\includegraphics[width=0.49\textwidth]{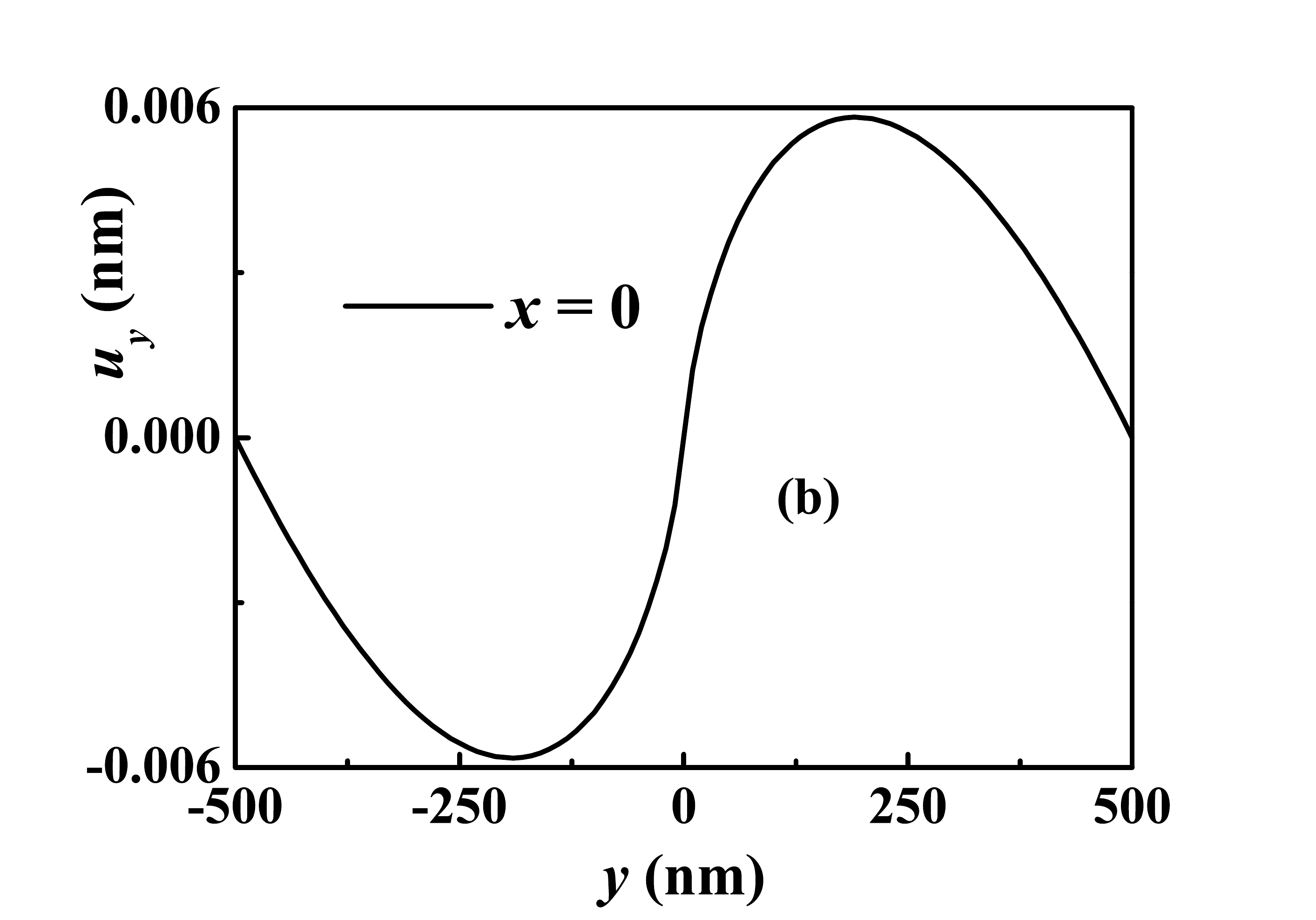}
	\caption{2D thin magnetic film. Steady state spatial profile of the displacement vector (a) $ u_x $ at $ y = 0 $, and (b) $ u_y $ at $ x = 0 $.}
	\label{fig_4}
\end{figure}

The reduction of the magnonic gap can be illustrated as follows: The
magnonic frequency $\omega\big(q,\vec{R}, \vec{\varepsilon}_{\xi \zeta}\big(\vec{R}\big)\big)$ increases with $q$.
Suppose the following equation holds:
$\omega\big(q_{1},\vec{R_{1}}, \vec{\varepsilon}_{\xi \zeta}\big(\vec{R_{1}}\big)\big)=\omega\big(q_{2},\vec{R_{2}}, \vec{\varepsilon}_{\xi \zeta}\big(\vec{R_{2}}\big)\big)$
for $q_{1}>q_{2}$. This means, in the vicinity of $\vec{R_{1}}$ the magnetoelastic coupling degrades the  magnonic frequency, or around $\vec{R_{2}}$ increases it. Thus, by
the constraint $\omega\big(q,\vec{R}, \vec{\varepsilon}_{\xi \zeta}\big(\vec{R}\big)\big)=const$ we can explore  the function $q\big(\vec{R}\big)$ or its inverse function.
Using the exponential temperature profile and the analytically derived deformation tensor $\vec{\varepsilon}_{\xi \zeta}\big(\vec{R}\big)$, the dispersion relation is calculated based on Eq. (\ref{magnon phonon modes}) with $ H_0 = 3 \times 10^5 $ A/m.  For a fixed frequency $ f = \omega / (2\pi) = 15 $ GHz, the profile of $ q(x,y) $ is shown in Fig. \ref{dispersion relation}(a). Similar to the temperature profile $ T(x, y) $ the symmetry features of the magnon profile manifests an isotropy in $xy$ plane. In the center ($ x = 0, y = 0 $), $ q $ reaches a minimum. The value of $ q $ increases gradually  with  distance from the center reaching a maximum to decrease near to the boundary. Since $\omega\big(q,\vec{R}, \vec{\varepsilon}_{\xi \zeta}\big(\vec{R}\big)\big)$ is fixed, an increase of the wave vector  is compensated by a negative contribution of the deformation tensor $\vec{\varepsilon}_{\xi \zeta}\big(\vec{R}\big)$ in the magnon dispersion relation. Therefore, the maximum of  $ q $ for a given fixed frequency $\omega\big(q,\vec{R}, \vec{\varepsilon}_{\xi \zeta}\big(\vec{R}\big)\big)$ corresponds to a minimum in the magnonic gap. We further calculate the elastic shift of the dispersion relations for different values of the coordinate $ x $ and a fixed value of the $ y = 0 $ coordinate, as shown in Fig. \ref{dispersion relation}(b). As we see, the magneto-elastic effect  can either increase the magnonic gap (Fig. \ref{dispersion relation}(b)) or may decrease depending on the geometry of the sample and on the parameters. We note that the value of the  gap is a local quantity that depends on  $\vec{R}$.

\section{Thermoelastic dispersion relations in thin magnetic films}

Having  explored the 3D  case of a bulk system we derive the thermoelastic dispersion relations for a  thin 2D magnetic film.
The solution of the elasticity equation for the displacement vector reads

\begin{eqnarray}
&&\vec{u}(\vec{R})=\chi\int(T(\vec{R}_1)-T_{0})\frac{\vec{R}-\vec{R}_1}{(R-R_1)^2}d^2\vec{R}_{1},\\
&&\chi=\frac{\kappa(1+\sigma)}{6\pi(1-\sigma)}.\nonumber
\end{eqnarray}
Similar to the 3D case, for 2D thin film we consider a  point-like and an extended heat source.

\subsection{Point-like heat source}

In particular for the point-like heat source $T(\vec{R}_1)-T_{0}=\frac{Q}{C}\delta(\vec{R}_1)$ we infer
\begin{equation}\label{Deformation2D}
\vec{u}(\vec{R})=\chi\frac{Q\vec{R}}{CR^2},
\end{equation}
while for the deformation tensor we obtain
\begin{equation}
\varepsilon_{\xi\zeta}=\frac{\chi Q}{C}\bigg(\frac{\delta_{\xi\zeta}}{R^2}-\frac{2 x_{\xi}x_{\zeta}}{R^4}\bigg).
\label{deformation tensor2D}
\end{equation}
We note that the plane deformation tensor $\varepsilon_{\xi\zeta}$ has three independent components: $\varepsilon_{xx},\varepsilon_{yy},\varepsilon_{xy}=\varepsilon_{yx}$. \\

We already see the difference to the bulk system. Instead of $1/|\vec{R}^{2}|$ for the bulk system (Eq.(\ref{1solution})), for the 2D thin magnetic film the
displacement vector decays slower  $1/|\vec{R}|$. The same applies to the deformation tensor Eq.(\ref{deformation tensor2D}).

The  magnetoacoustic energy density of the thin film has the form
 \begin{eqnarray}\label{magnitacousticenergy2D}
&& U_{mel}(R)=\\
&&=\frac{B_{1}}{M_{s}^2}\big(M_{x}^{2}\varepsilon_{xx}+M_{y}^{2}\varepsilon_{yy}\big)+\frac{2B_{2}}{M_{s}^2}M_{x}M_{y}\varepsilon_{xy},\nonumber
 \end{eqnarray}
and the effective  magnetoacoustic field is
 \begin{eqnarray}\label{Heff2d}
 H_{xeff}=-\frac{2B_{1}}{M_{s}^2}\varepsilon_{xx}M_{x}-\frac{2B_{2}}{M_{s}^2}\varepsilon_{xy}M_{y},\\
  H_{yeff}=-\frac{2B_{2}}{M_{s}^2}\varepsilon_{yx}M_{x}-\frac{2B_{1}}{M_{s}^2}\varepsilon_{xx}M_{x}.\nonumber
 \end{eqnarray}
Utilizing  Eq.({\ref{LLG}}), Eq.(\ref{magnitacousticenergy2D}), and Eq.({\ref{Heff2d}}) and assuming that
the ground state magnetization is aligned parallel to the
$z$ axis we derive the following dispersion relation of the
coupled magnetoelastic magnonic modes in the thin films as

\begin{eqnarray}\label{Dispersion2D}
&&\omega(q,R)=\frac{1}{\hbar}\sqrt{\big(Aq^2+g^{(1)}_{D}\big)\big(Aq^2+g^{(2)}_{D}\big)-g^{(3)}_{D}},\nonumber\\
&&g^{(1)}_{D}=\hbar\gamma\big(H_{0}+\frac{2B_{1}}{M_{s}^2}\varepsilon_{xx}\big),\\
&&g^{(2)}_{D}=\hbar\gamma\big(H_{0}+\frac{2B_{1}}{M_{s}^2}\varepsilon_{yy}\big),\nonumber\\
&&g^{(3)}_{D}=\Big(\hbar\gamma\frac{2B_{2}}{M_{s}^2}\varepsilon_{xy}\Big)^2.\nonumber
\end{eqnarray}
As we see from Eq.(\ref{Dispersion2D}) the dispersion relation is defined by the external field $H_{0}$ and the deformation tensor $\varepsilon_{\xi\zeta}$.

\subsection{Extended heat source}

In order to explore the effect of the extended heat source we solve the heat equation (\ref{heat equation}):
\begin{eqnarray}\label{heatequationstationary}
\frac{\partial^2T(x,y)}{\partial x^2}+\frac{\partial^2 T(x,y)}{\partial y^2}=-aI(x,y),a=\frac{\rho C}{k_{ph}}
\end{eqnarray}
We adopt the source term $I=I_{0}e^{-\alpha(x+y)}$, with a  positive characteristic decay constant $\alpha>0$ and the following boundary conditions: $x>0, y>0$.
Then, the stationary solution of Eq.(\ref{heatequationstationary}) reads

\begin{eqnarray}\label{solutionHeat2D}
&&T(x,y)=-\frac{a I_{0}}{2 \alpha^2}e^{-\alpha(x+y)}+T_{0},\nonumber\\
&&\alpha>0,  x>0, y>0.
\end{eqnarray}
Taking into account Eq.(\ref{solutionHeat2D}) for the displacement vector and the deformation tensor we deduce following solutions
\begin{eqnarray}\label{deformation2DExtended}
&&u_{x}(x,y)=\\
&&=-\chi\frac{aI_{0}}{2 \alpha^2}\int_{0}^{\infty}\int_{0}^{\infty}e^{-\alpha(x_1+y_1)}\frac{(x-x_{1})dx_1dy_1}{(R-R_{1})^2}\nonumber \\
&&u_{y}(x,y)=\nonumber \\
&&=-\chi\frac{aI_{0}}{2 \alpha^2}\int_{0}^{\infty}\int_{0}^{\infty}e^{-\alpha(x_1+y_1)}\frac{(y-y_{1})dx_1dy_1}{(R-R_{1})^2}\nonumber
\end{eqnarray}
and
\begin{eqnarray}
&&\varepsilon_{xx}=-\varepsilon_{yy}=\\
&&=\chi\frac{aI_{0}}{2 \alpha^2}\int_{0}^{\infty}\int_{0}^{\infty}e^{-\alpha(x_1+y_1)}\frac{(x-x_{1})^2-(y-y_{1})^2}{(R-R_{1})^4}dx_1dy_1,\nonumber\\
&&\varepsilon_{xy}=\varepsilon_{yx}=\nonumber \\
&&=\chi\frac{aI_{0}}{\alpha^2}\int_{0}^{\infty}\int_{0}^{\infty}e^{-\alpha(x_1+y_1)}\frac{(x-x_{1})(y-y_{1})}{(R-R_{1})^4}dx_1dy_1,\nonumber\\
&&(R-R_1)^2=(x-x_1)^2+(y-y_1)^2.\nonumber
\end{eqnarray}

The obtained result is quite involved  in general and can be made transparent by numerical simulations. In the isotropic case, the
problem simplifies and we obtain an analytical solution in a closed form.

\subsection{Extended isotropic heat source}
We assume that source term is isotropic $I=I_0e^{-\alpha R}$ and the temperature is a function of $R$ only.
Utilizing polar coordinates and performing the integration over angle we obtain from
 Eq.(\ref{heatequationstationary})
\begin{equation}
\frac{1}{R}\frac{\partial}{\partial R}\bigg(R\frac{\partial T(R)}{\partial R}\bigg)=-aI(R).
\end{equation}
We adopt the boundary condition $T(R\rightarrow \infty)=T_{0}$ and find the  stationary solution of the heat equation in the following form:
\begin{equation}\label{solutionheatisotropic}
T(R)-T_{0}=-\frac{I_0}{\alpha^2}\big(\Gamma(0,\alpha R)+e^{-\alpha R}\big).
\end{equation}
Here $\Gamma(\alpha,z)=\int_{z}^{\infty}t^{\alpha-1}e^{-t}dt$ is the incomplete Gamma function.

Taking into account the temperature profile Eq.(\ref{solutionheatisotropic}), for the displacement vector and the deformation tensor we deduce the following solutions
\begin{eqnarray}
&&u_{x}=-\chi\frac{I_0}{\alpha^2}\int_{0}^{\infty}\int_{0}^{2\pi}\big(\Gamma(0,\alpha R')+e^{-\alpha R'}\big)\times\\
&&\frac{R \cos \theta-R' \cos \theta'}{R^2+R'^{2}-2RR'\cos(\theta-\theta')}R'dR'd\theta',\nonumber \\
&&u_{y}=-\chi\frac{I_0}{\alpha^2}\int_{0}^{\infty}\int_{0}^{2\pi}\big(\Gamma(0,\alpha R')+e^{-\alpha R'}\big)\times\\
&&\frac{R \sin \theta-R' \sin \theta'}{R^2+R'^{2}-2RR'\cos(\theta-\theta')}R'dR'd\theta'.\nonumber
\end{eqnarray}
After performing the integration
for the displacement vector and the deformation tensor we arrive at the expression
\begin{eqnarray}\label{Deformation2Disotropic}
u_{x}=-\chi\frac{3\pi I_{0}}{2\alpha^4}\frac{x}{R^2},\\
u_{y}=-\chi\frac{3\pi I_{0}}{2\alpha^4}\frac{y}{R^2},
\end{eqnarray}
and
\begin{eqnarray}\label{deformation tensor 2D isotropic}
&&\varepsilon_{xx}=-\varepsilon_{yy}=\chi\frac{3\pi I_{0}}{2\alpha^4}\frac{x^2-y^2}{R^4},\\
&&\varepsilon_{xy}=\varepsilon_{yx}=\chi\frac{3\pi I_{0}}{\alpha^4}\frac{xy}{R^4}.
\end{eqnarray}
Here $R^2=x^2+y^2$ is the in-plane radius vector. In order to obtain the dispersion relations in a closed form, we use Eq.(\ref{Deformation2Disotropic}) - Eq.(\ref{deformation tensor 2D isotropic})
in  Eq.(\ref{Dispersion2D}). Plots of the displacement vector, the deformation tensor and the dispersion relation
in the case of a 2D extended heat source are presented in Fig. \ref{fig_3-2d}. As we see the magnon dispersion relation is  local
and is different in the different areas  of the film.

\subsection{Linear temperature profile in the thin magnetic films}

We again consider the linear temperature profile of the following form $T(R)=-\big(T'-T_{0}\big)R/R_{max}+T'$, where $T'>T_{0}$, and the temperature at the edges is equal to $T(0)=T'$,
$T(R_{max})=T_{0}$. After implementing liner temperature profile, for displacement vector we deduce:

\begin{eqnarray}\label{Deformation2Linear}
u_{x}=-\chi(T_{0}-T')\frac{\pi R_{max}^2}{6}\frac{x}{R^2},\\
u_{y}=-\chi(T_{0}-T')\frac{\pi R_{max}^2}{6}\frac{y}{R^2},
\end{eqnarray}

\textbf{and for the deformation tensor:}

\begin{eqnarray}\label{deformation tensor 2D Linear}
&&\varepsilon_{xx}=-\varepsilon_{yy}=\chi(T_{0}-T')\frac{\pi R_{max}^2}{6}\frac{x^2-y^2}{R^4},\\
&&\varepsilon_{xy}=\varepsilon_{yx}=\chi(T_{0}-T')\frac{\pi R_{max}^2}{6}\frac{xy}{R^4}.
\end{eqnarray}

As we see linear temperature profile in the thin magnetic films lead to the same type of asymptotic decay, $1/R$ for dismastment vector and $1/R^{2}$ for
the deformation tensor.

\section{numerical results}
The numerical simulations are performed for a two dimensional Nickel film.
The external field with $ H_z = 4.5 \times 10^5 $ A/m is applied along the \textit{x} axis leading to a uniform ground state. A two-dimensional ferromagnetic thin film with a  length of 1000 nm is aligned along \textit{x} axis and has a width of 1000 nm in \textit{y} direction. We also assume that the source term that enters in the heat equation and describes the effect of the laser heating is a local function $I\big(\vec{R}, t\big)=\delta\big(\vec{R}-\vec{R}_{0}\big)$ corresponding to the case when the  intensity of the laser field is spatially localized. This approximation is similar to the local heat source model discussed analytically in the previous sections.
We assume the heating  laser spot is around  ($ x = 0 $ nm, $ y = 0 $ nm). The thermal effect of the laser heating is described by the heat equation Eq. (\ref{heat equation}) using the fixed boundary conditions
\begin{equation}
\begin{aligned}
\displaystyle
T(x = 500 \mathrm{nm}) = 0, \\
T(x = -500 \mathrm{nm}) = 0, \\
T(y = 500 \mathrm{nm}) = 0, \\
T(y = -500 \mathrm{nm}) = 0.
\label{eq_1}
\end{aligned}
\end{equation}

\begin{figure}
	\centering
	\includegraphics[width=0.49\textwidth]{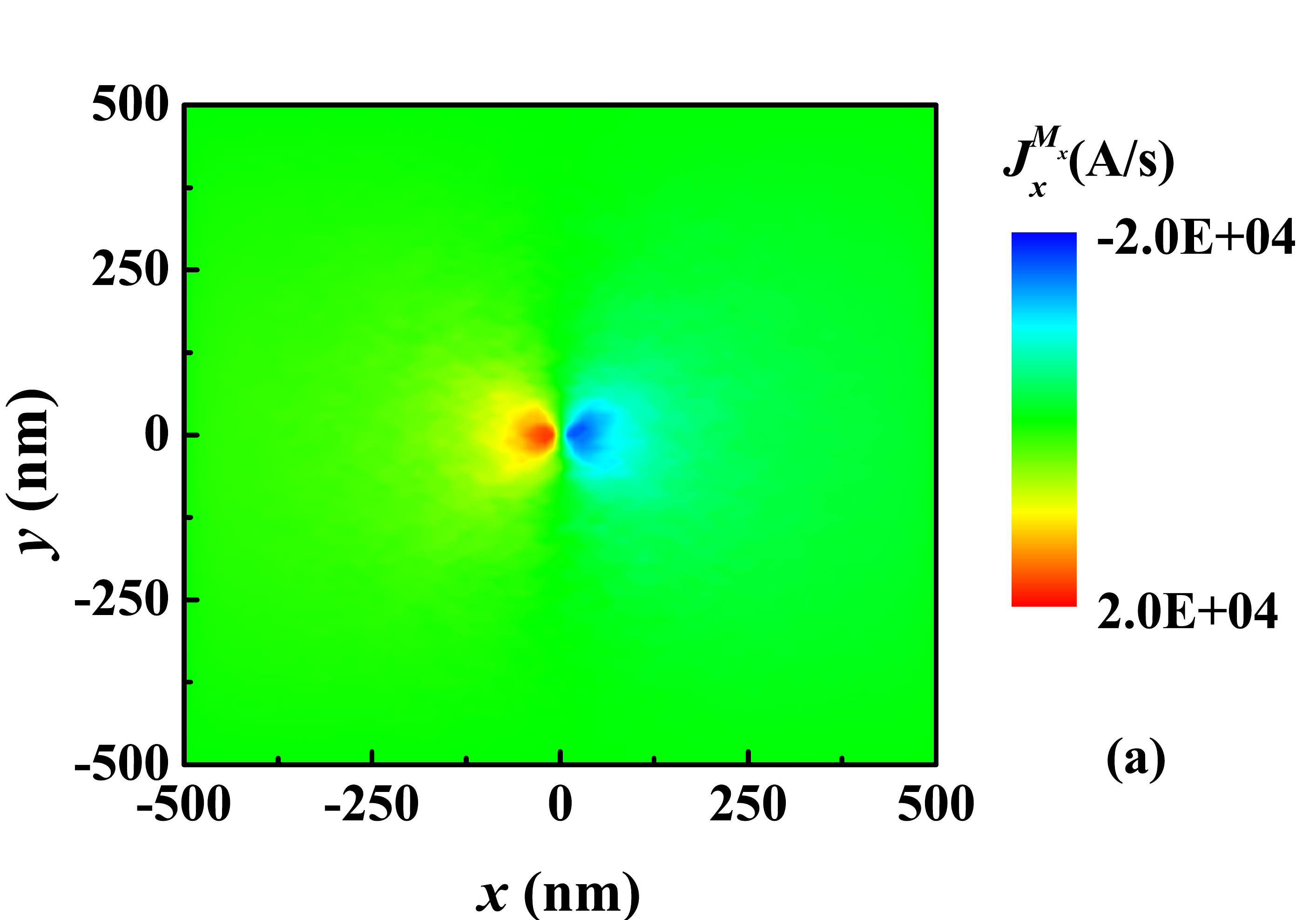}	
	\includegraphics[width=0.49\textwidth]{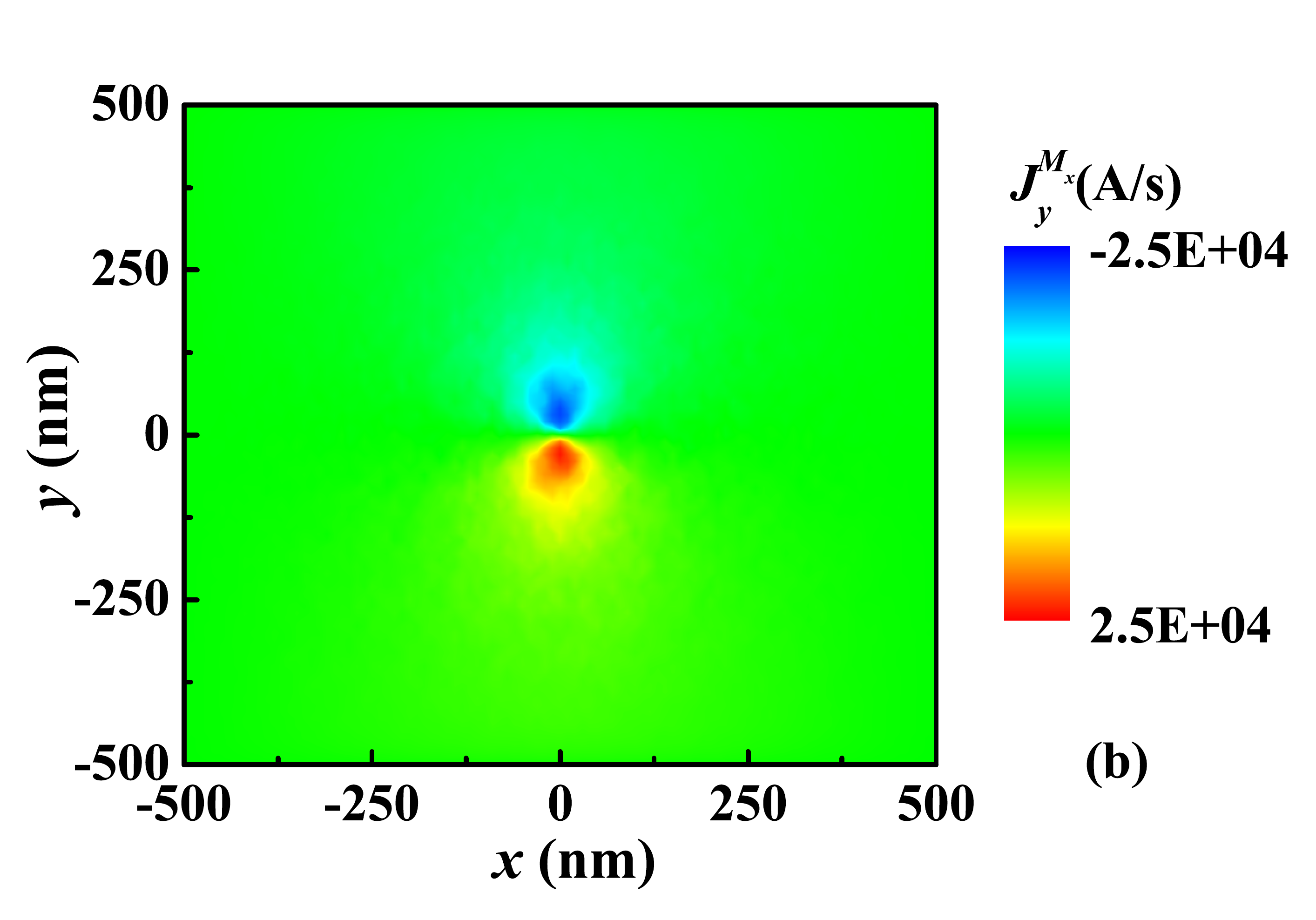}	
	\caption{2D thin magnetic film. The steady state spatial profile of the components of the magnonic spin current tensor (a) $ J_x^{M_x} $, and (b) $ J_y^{M_x} $ with an elastic term when $ T_{0} = 80 $K. }
	\label{fig_5}
\end{figure}

\begin{figure}
	\centering
	\includegraphics[width=0.49\textwidth]{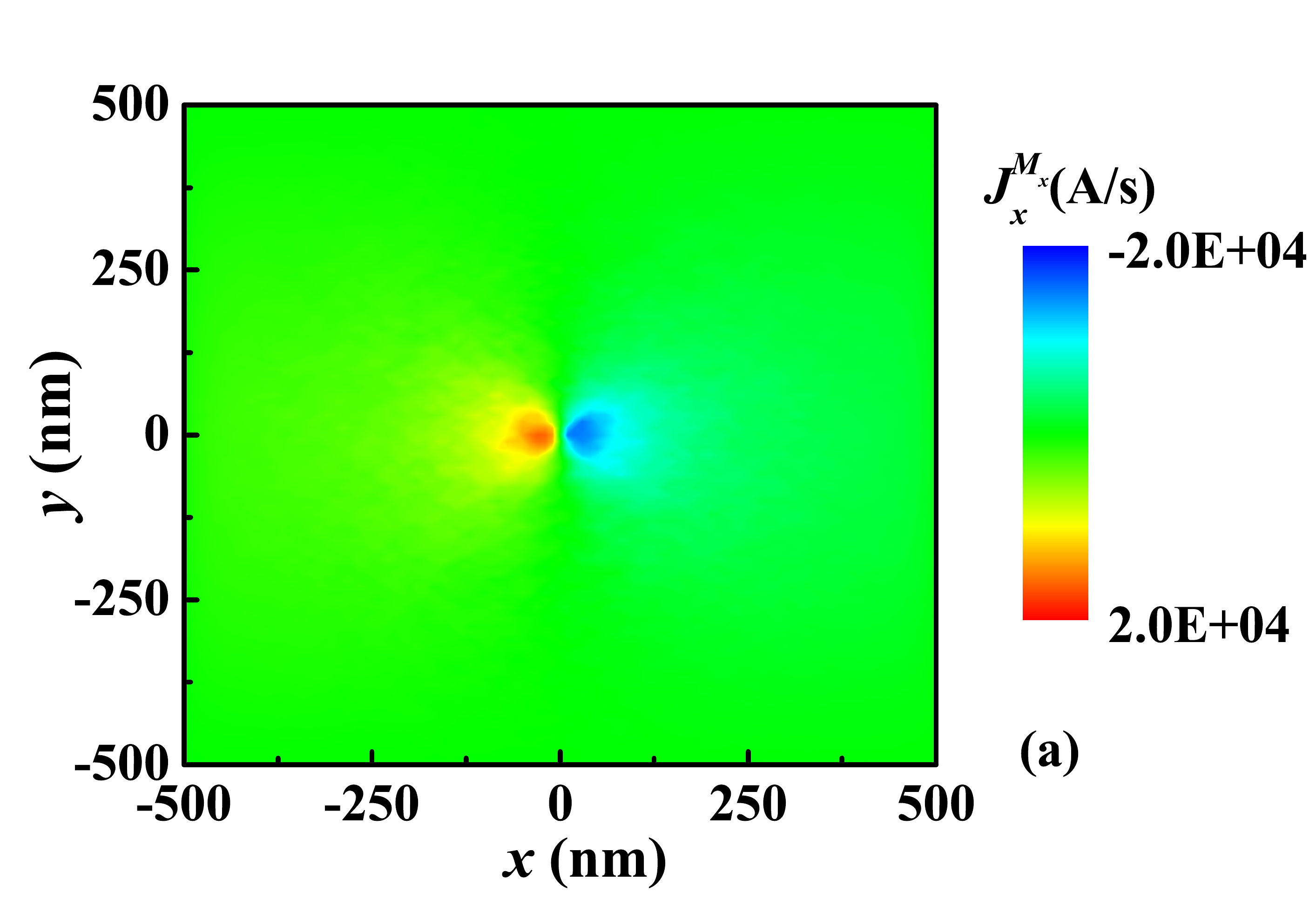}	
	\includegraphics[width=0.49\textwidth]{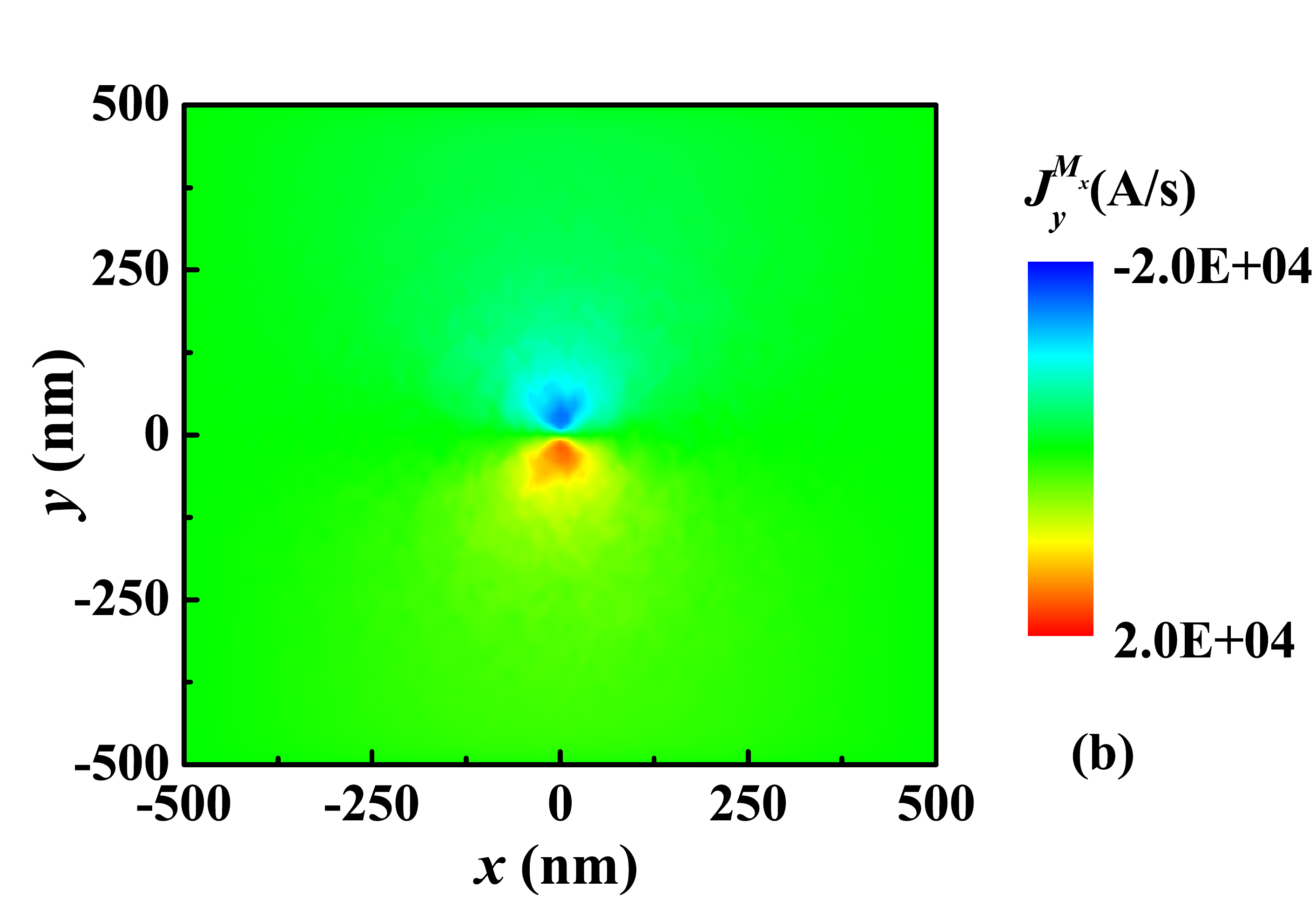}	
	\caption{2D thin magnetic film. The steady state spatial profile of the components of the magnonic spin current tensor in the absence of a magneto-elastic effect (a) $ J_x^{M_x} $, and (b) $ J_y^{M_x} $ when $ T_{0} = 80 $K. }
	\label{fig_6}
\end{figure}

\begin{figure}
	\centering
	\includegraphics[width=0.49\textwidth]{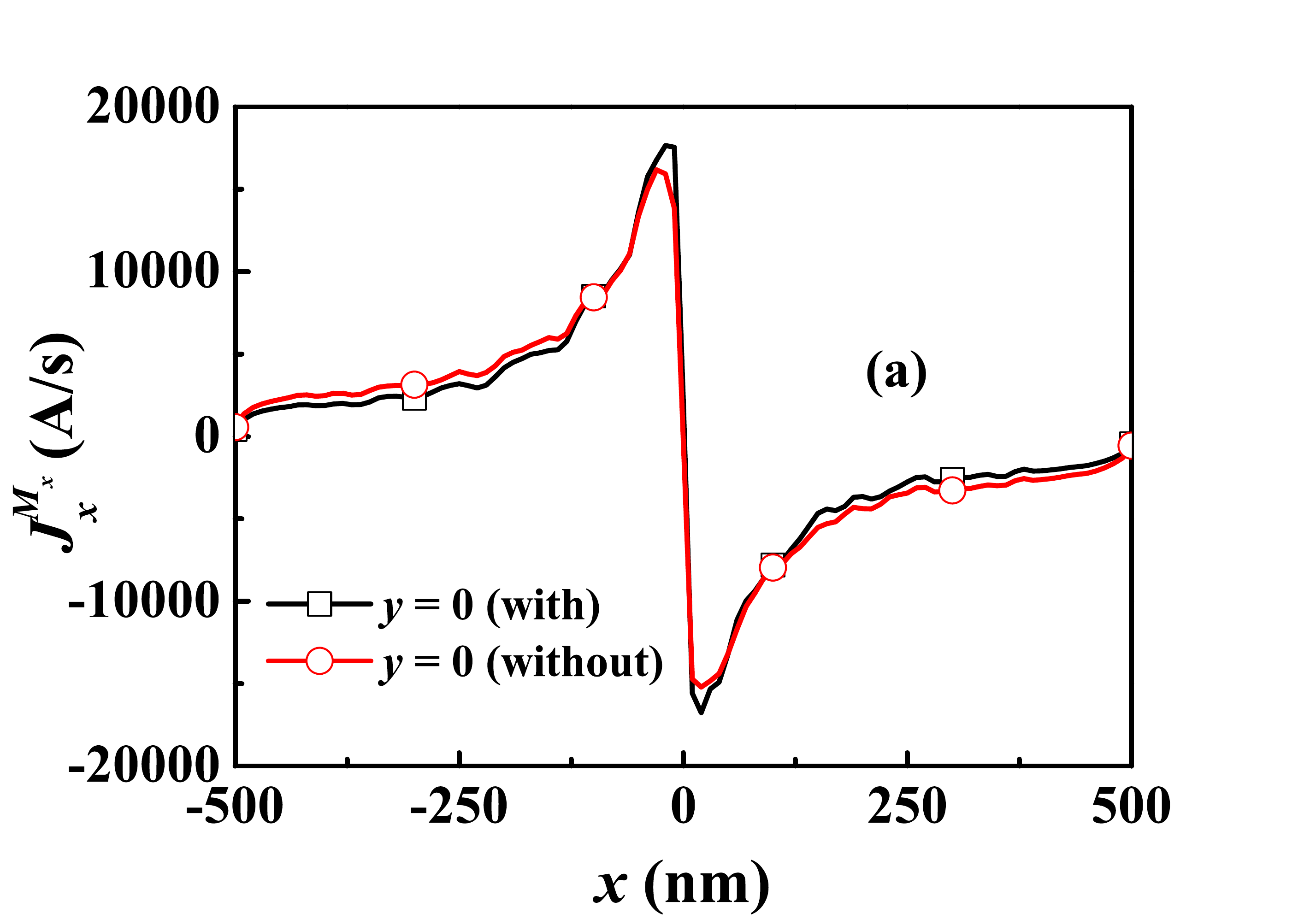}	
	\includegraphics[width=0.49\textwidth]{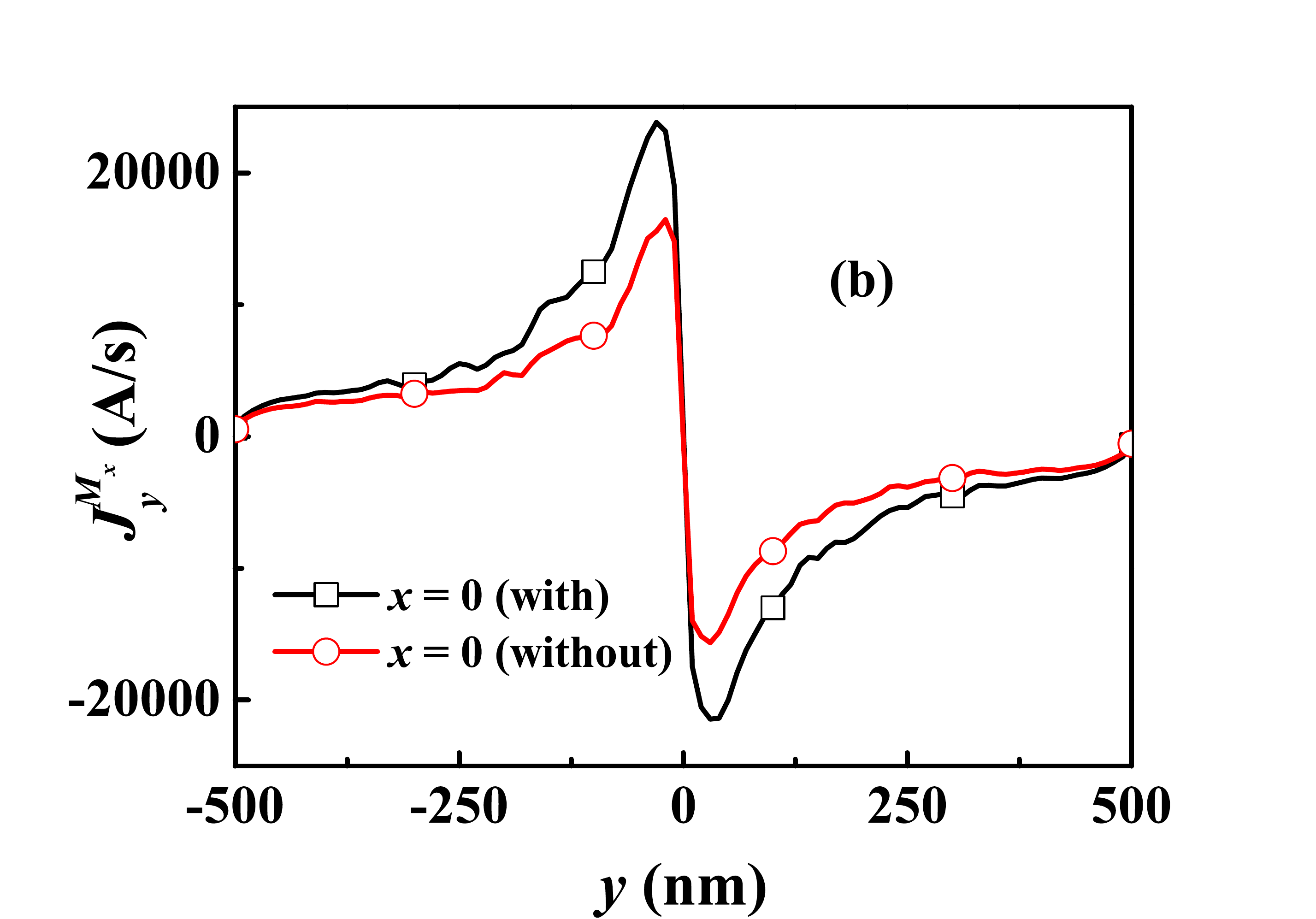}	
	\caption{2D thin magnetic film. The steady state spatial profile of the components of magnonic spin current tensor (a) $ J_x^{M_x} $ at $ y = 0 $, and (b) $ J_y^{M_x} $ at $ x = 0 $ with and without the magneto-elastic term.}
	\label{fig_7}
\end{figure}

The value of the temperature $ T_0 $ formed in the area of the heating laser  depends on the laser intensity $ I $. For the boundary conditions Eq.(\ref{eq_1}) implemented numerically in the heat equation Eq. (\ref{heat equation})  we obtain the stable spatial temperature  profile. As  inferred  from  Fig. \ref{fig_1} , the temperature profile $ T(x, y) $ shows an exponential decay with the distance from $T_{0}\big(x=0,~y=0\big)$. This numerical result for the temperature profile is in a good agreement with the assumptions used for the analytical solution.  The exponential character of the decay of the temperature profile is more evident from   Fig. \ref{fig_2}.

The heat diffusion induces a thermal gradient $ \nabla T $ and an elastic deformation in the system. Using the equation of elasticity Eq. (\ref{thermal elastic wave})  we  find numerically the temperature profile $ T(x, y) $ and  calculate the components of the displacement vector for the following fixed boundary conditions
\begin{equation}
\begin{aligned}
\displaystyle \vec{u}(x = 500 \mathrm{nm}) = (0, 0, 0), \\
\vec{u}(x = -500 \mathrm{nm}) = (0, 0, 0), \\
\vec{u}(y = -500 \mathrm{nm}) = (0, 0, 0), \\
\vec{u}(y = 500 \mathrm{nm}) = (0, 0, 0).
\label{eq_2}
\end{aligned}
\end{equation}

The steady state components of the elastic displacement vector $ u_x $ and $ u_y $ are shown in Fig. \ref{fig_3}, and Fig. \ref{fig_4}.
We recognize  a certain similarity with the previously obtained analytical results. Namely, we clearly see that the $u_{x}$ component exhibits the mirror symmetry with respect to the reflection $y\rightarrow - y$, and is antisymmetric with respect to the reflection $x\rightarrow - x$. Concerning the component $ u_y $, the  situation is reversed: symmetry is given for $x\rightarrow - x$ and antisymmetry for  $y\rightarrow - y$. The components of the deformation tensor $ \epsilon_{xx} $, $ \epsilon_{yy} $ and $ \epsilon_{xy} $ are shown in Fig. \ref{eps-numxx}, Fig. \ref{eps-numyy}, Fig. \ref{eps-numxy}. The diagonal components $ \epsilon_{xx} $ and $ \epsilon_{yy} $ are larger but localized, while  the non-diagonal component of the deformation tensor $ \epsilon_{xy} $ decays slower   and remains  finite in the whole sample.

We implement the numerically deduced  temperature profile $ T(x, y) $ and the elastic displacement profile $ \vec{u}(x, y) $, in the stochastic LLG equation Eq. (\ref{LLG}). The existence of the thermal gradient formed in the system  due to the laser heating may lead to the emergence of magnonic spin current and a longitudinal spin Seebeck effect. However,  on top of this standard effect, the thermal heating leads to a thermal activation of the deformation tensor. Due to the magneto-elastic interaction the thermally activated deformation tensor contributes to the magnetization dynamics and modifies the net magnonic current. For, $ T_{0} = 80 $K, the components of the magnonic spin current tensor $ J_x^{M_x} = \frac{2 \gamma A_{ex} }{\mu_0 M_s^2} (M_y \partial_x M_z - M_z \partial_x M_y) $ and $ J_y^{M_x} = \frac{2 \gamma A_{ex} }{\mu_0 M_s^2} (M_y \partial_y M_z - M_z \partial_y M_y) $ are plotted in Fig. \ref{fig_5}.

\begin{figure}
	\centering
	\includegraphics[width=0.49\textwidth]{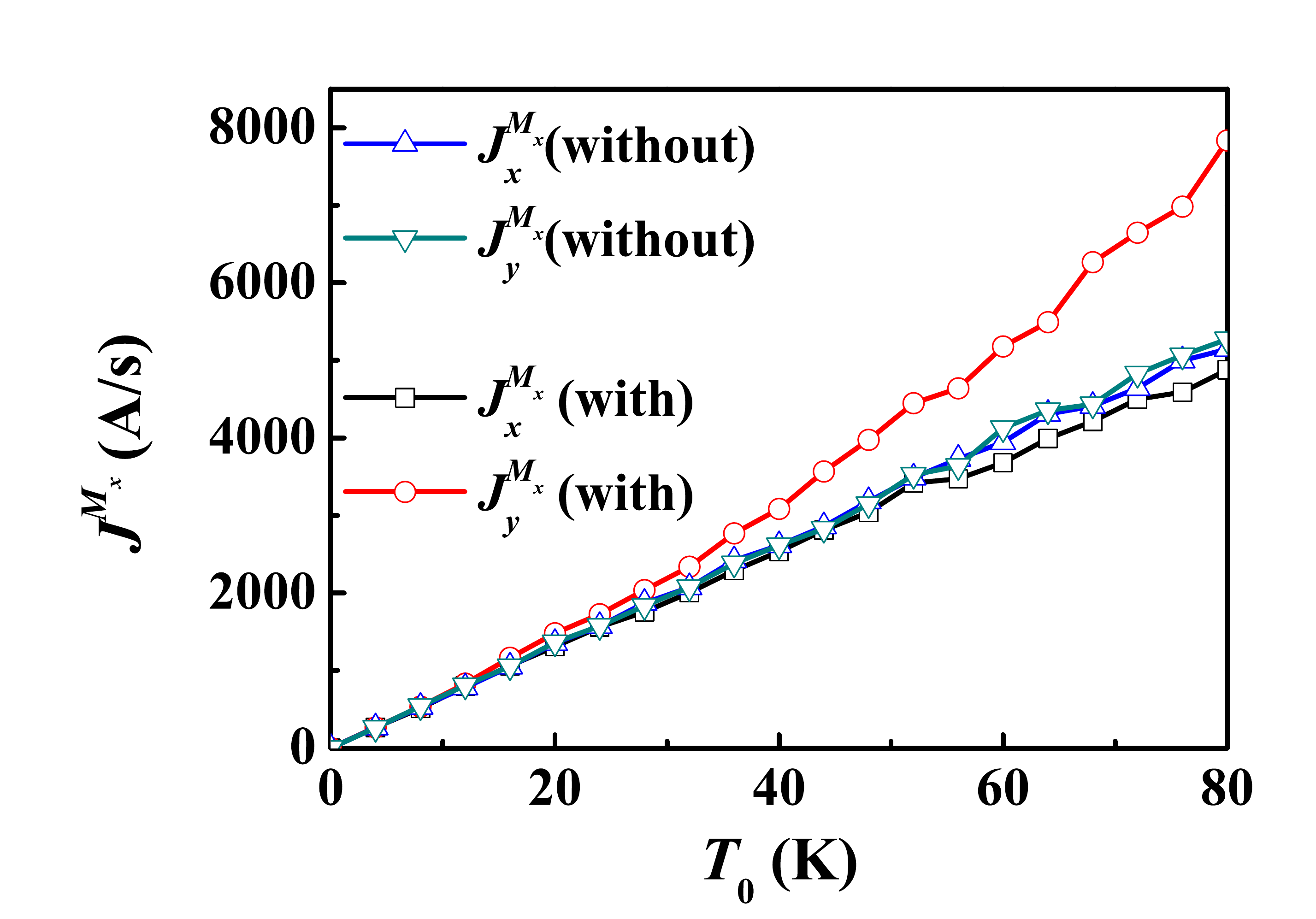}	
	\caption{2D thin magnetic film. The averaged $ J_x^{M_x} $ in  $ ((-500) \mathrm{nm} < x < 0, y = 0) $ and $ J_y^{M_x} $ in $ (x = 0, (-500) \mathrm{nm} < y < 0) $ with and without the magneto-elastic term.}
	\label{zong-xy-spincurrent(Mx)}
\end{figure}

\begin{figure}
	\centering
	\includegraphics[width=0.49\textwidth]{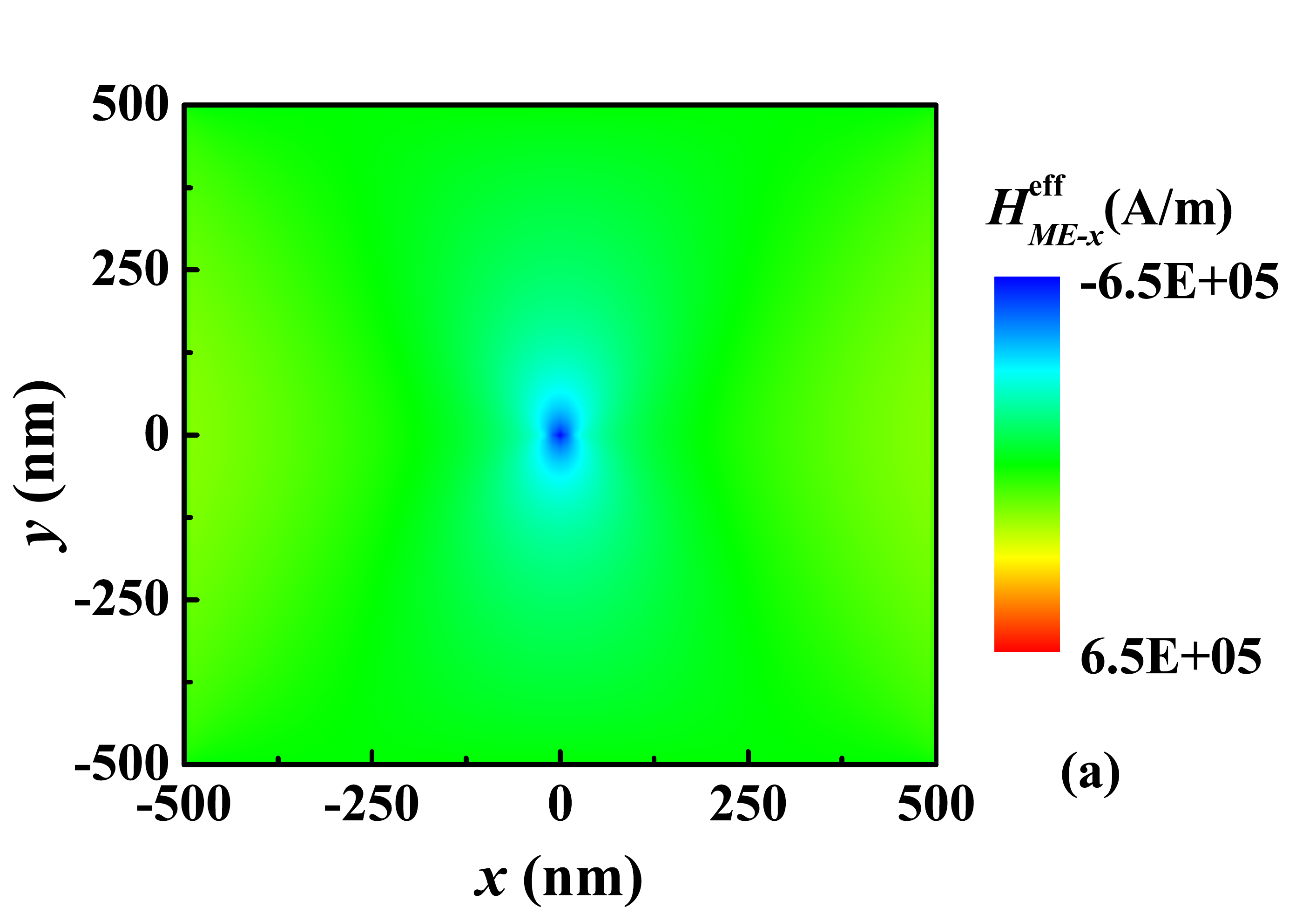}	
	\includegraphics[width=0.49\textwidth]{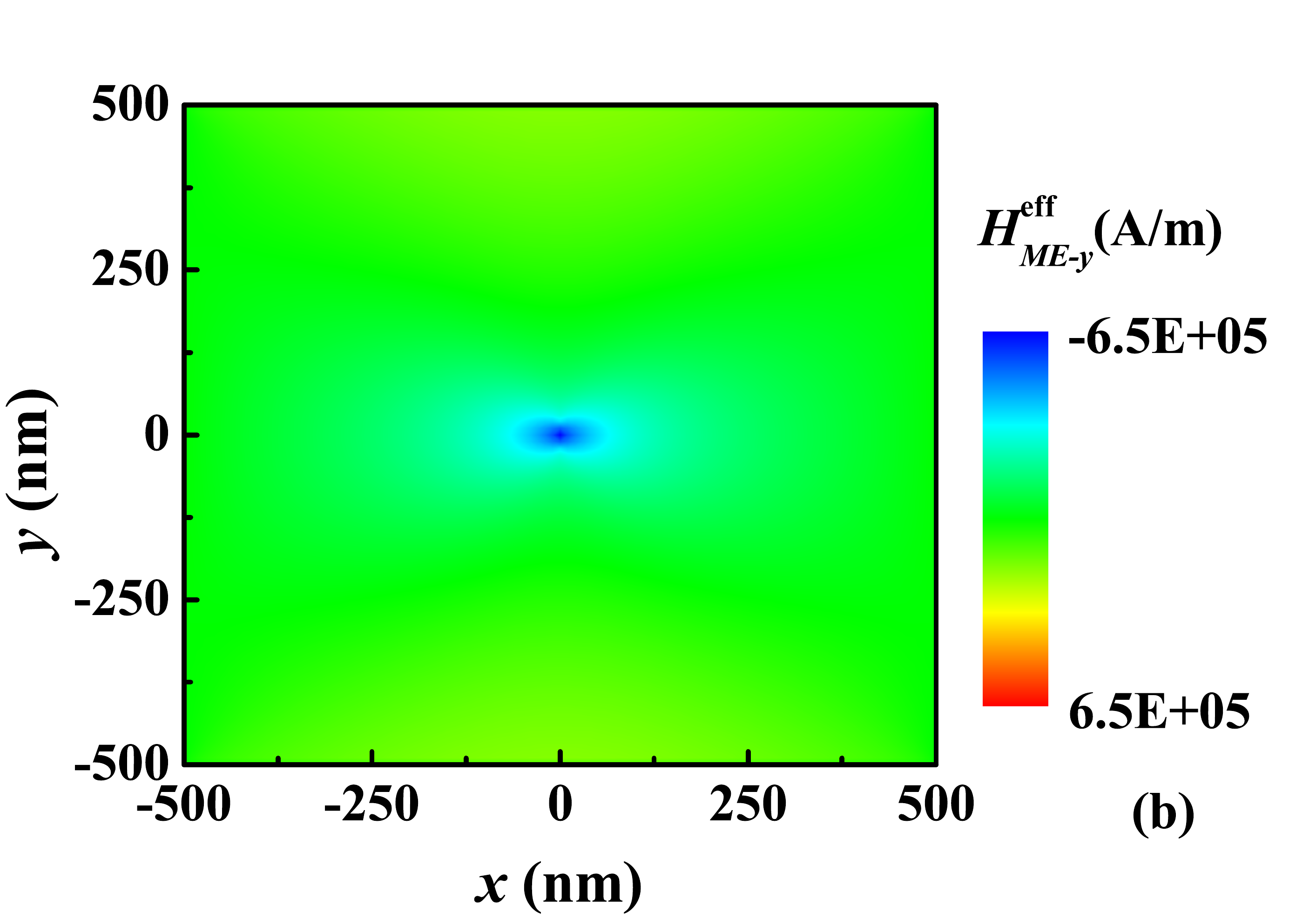}		
	\caption{2D thin magnetic film. (a) The profile of the numerically obtained  \textit{x}-component of the magneto-elastic effective field $ H^{\mathrm{eff}}_{ME-x} $ with the equilibrium magnetization in the $ +x $ direction. (b) The profile of the  \textit{y}-component of the magneto-elastic effective field $ H^{\mathrm{eff}}_{ME-y} $, calculated numerically  with the equilibrium magnetization in the $ +y $ direction.}
	\label{H-me-x-Mx}
\end{figure}

 Fig. \ref{fig_6} evidences that the magnon current emerges in the area heated by the laser pulses and propagates according to the formed thermal bias (which has been established swiftly on the magnon time scale). The $ J_x^{M_x} $ and $ J_y^{M_x} $ components of the spin current tensor are negative for $ x >0 $ and $ y>0 $, respectively becoming  positive in the rest of the sample. A similar abrupt switching of the displacement vector  $ u_x $ and $ u_y $ we observe in Fig. \ref{fig_3}(b).  To highlight  the role of  the elastic term on the magnonic spin current we plot the magnonic current in the presence/absence of the magneto-elastic contribution for the same thermal gradient.
Fig. \ref{fig_6} shows the profile of the magnonic spin current, in particular the tensor components $ J_x^{M_x} $ and $ J_y^{M_x} $. They have qualitatively  the same behavior in the presence or in the absence of the magneto-elastic coupling. However, as shown in Fig \ref{fig_7}, the value of $ J_y^{M_x} $ is obviously becoming larger with  the elastic term, while the change in  $ J_x^{M_x} $  due to the elastic term  is marginal.
We conclude so that elasticity leads to the enhancement of the magnonic spin Seebeck current in certain directions. Also,  the bias temperature is important. Raising  $ T_{0} $ and hence $ \vec{u} $, the increase  in the magnonic spin current $ J^{M_x} $, and the elastic enchantment of $ J_y^{M_x}  $ are evident, as shown in Fig. \ref{zong-xy-spincurrent(Mx)}.

\begin{figure}
	\centering	\includegraphics[width=0.49\textwidth]{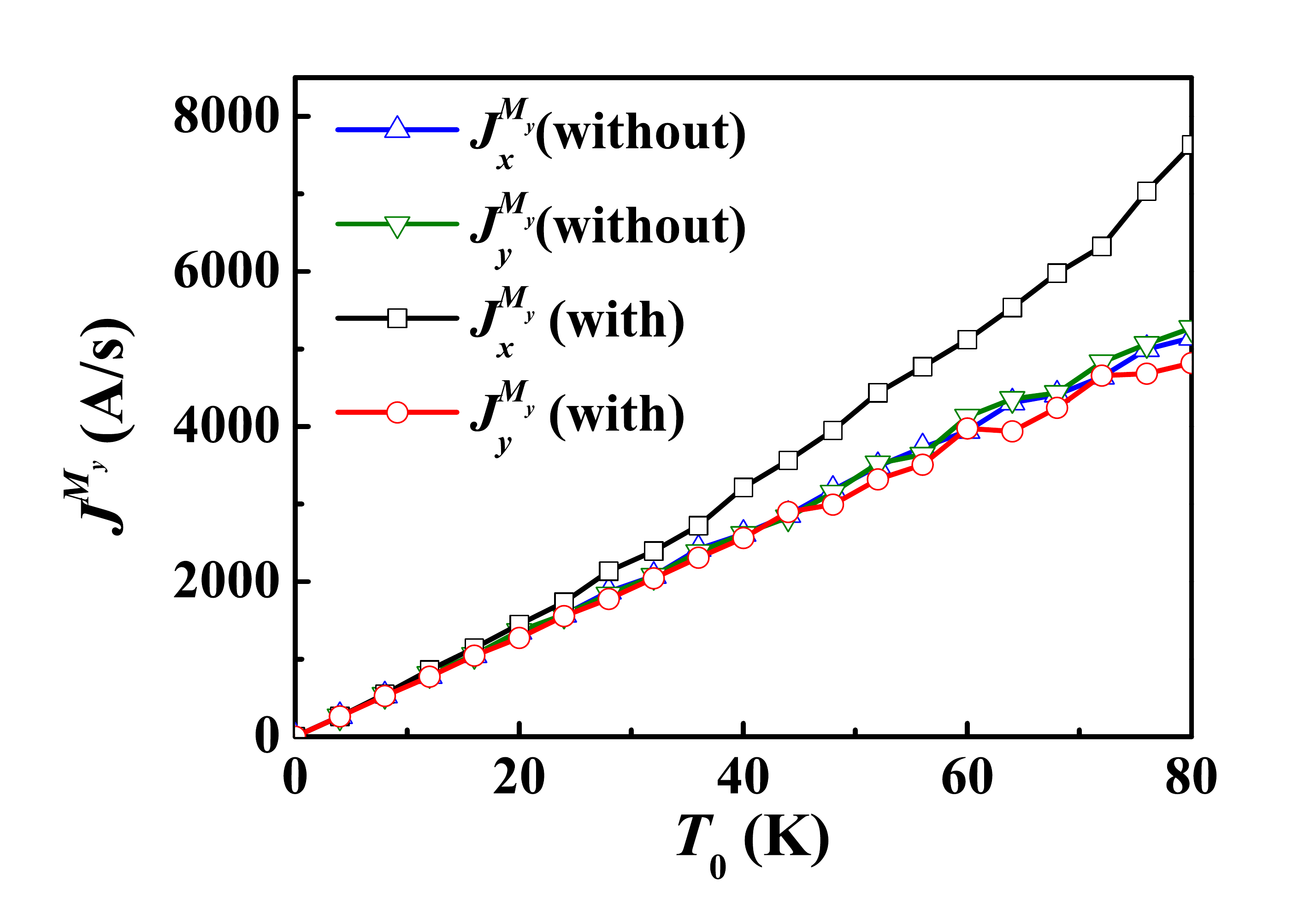}	
	\caption{2D thin magnetic film. The averaged $ J_x^{M_y} $ in  $ ((-500) \mathrm{nm} < x < 0, y = 0) $ and $ J_y^{M_y} $ in $ (x = 0, (-500) \mathrm{nm} < y < 0) $ with and without the magneto-elastic term.}
	\label{zong-xy-spincurrent(My)}
\end{figure}
To understand the selective enhancement of  the magnonic spin current induced by the magneto-elastic coupling
we inspect  the \textit{x}-component of the magneto-elastic effective field $ H^{\mathrm{eff}}_{ME-x}=-\big(\delta U_{mel}(\vec{R})/\delta \vec{\vec{M}}(\vec{R})\big)_{x}$ in Fig. \ref{H-me-x-Mx}(a) which shows that the effective magneto-elastic filed is directed opposite to the external field.  Therefore, it reduces the gap in the magnon spectrum. By changing the equilibrium magnetization along $ +y $, the distribution of the effective magneto-elastic filed is changed (Fig. \ref{H-me-x-Mx}(b)), and the corresponding magnonic spin current, i.e. $ J_x^{M_y} $, is selectively enhanced. This feature is further testified by Fig. \ref{zong-xy-spincurrent(My)}. The maximal temperature in the center of the laser intensity is equal to $ T_{0} = 80 $ K.  For the experimental observation for the thermoelastic effect, we suggest exploiting the magnetization sensitive thermoelastic effect, and detecting the spin Seebeck effect for the different direction of the equilibrium magnetization.

\begin{figure}
	\centering	\includegraphics[width=0.49\textwidth]{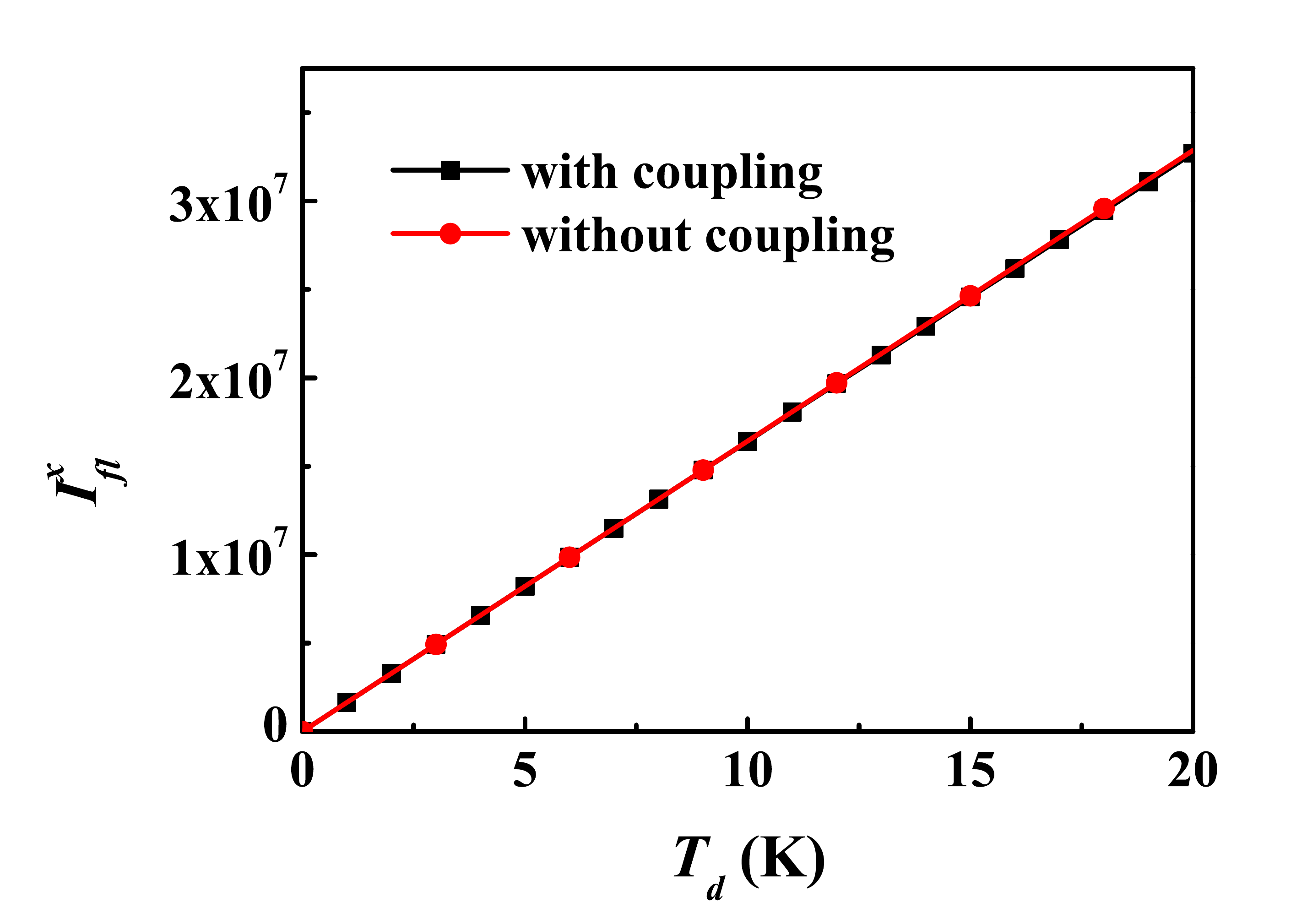}	
	\caption{2D thin magnetic film. The $  x $ component of the backward spin current $\vec{I}^x_{fl}$ in the presence and absence of the thermoelastic effect. The temperature $ T_d $ of the detecting bar is uniform, while the temperature of the magnet is inhomogeneous. The equilibrium magnetization is along $ +x $.}
	\label{backward current}
\end{figure}

At the typical Spin Seebeck experiment, spin current is measured by means of inverse spin Hall effect and Platinum detecting bar. The total net current $\vec{I}=\vec{I}_{sp}+\vec{I}_{fl}$ consists of two contributions:
spin pumping current $\vec{I}_{sp}$ flowing from magnetic insulator towards detecting bar, and a backward spin current $\vec{I}_{fl}=-M_{s}\vec{m}\times\vec{h}$ injected from the detecting bar into the magnetic insulator. \cite{Xiaonew}
Here $\vec{h}$ is the random magnetic field related to the Platinum detecting bar. The interesting question is whether the thermoelasticity plays an important role in the backward spin current.
In order to answer this question, we calculate the backward spin current $\vec{I}_{fl}$ in the presence and absence of the thermoelastic effect. The result of the numerical calculations is shown in Fig. \ref{backward current}. As we thermoelasticity see has no influence on the backward spin current.

\section{Conclusions}
We studied the spin Seebeck effect in bulk samples and  thin ferromagnetic films and explored the influence of the thermoelastic steady state deformation on the magnonic spin current.
For a particular temperature profile in the system, we obtained analytical expressions for the thermoelastic deformation tensor.
We derived analytical results for the 3D bulk system and 2D thin magnetic film as well. We observed that the displacement vector and the deformation tensor in bulk systems decay asymptotically
as $u\sim1/R^{2}$, and $\varepsilon\sim1/R^{3}$, respectively. The decay in  thin magnetic films is slower being $u\sim1/R$, and $\varepsilon\sim1/R^{2}$.
We found that due to the magnetoelastic coupling, the thermoelastic deformation tensor has a significant impact on the magnetization dynamics. We derived analytical expressions for the dispersion relations for thermoelastic magnons highlighting a principle difference between the thermoelastic and the magneto-elastic effects. Magnetoelastic effects always enhance the magnetoelastic gap in the magnonic spectrum.
Thermoelastic steady state deformation may lead  either to an enchantment or to a reduction in the gap of the magnonic spectrum.
A reduction of the gap increases the number of magnons contributing to the spin Seebeck effect
offering so  a thermoelastic control of the spin Seebeck effect.

\section{Acknowledgment} This work is supported by the DFG through the SFB 762, SFB-TRR 227, as well as by National Natural Science Foundation of China (No. 11704415).


\end{document}